\documentclass[10pt,twocolumn]{article}
\usepackage[utf8]{inputenc}
\usepackage[english]{babel}
\usepackage[a4paper, margin=0.75in]{geometry}
\usepackage{graphicx}
\usepackage{float}
\usepackage{authblk}
\usepackage{lipsum}
\usepackage{amsmath,amsfonts,amssymb}
\usepackage{physics}
\usepackage{soul}
\usepackage{color}
\usepackage{titlesec}
\usepackage{chngcntr}
\usepackage{enumitem} %
\usepackage{mathtools} %
\usepackage{csquotes}

\RequirePackage[normalem]{ulem} %
\RequirePackage{color}\definecolor{RED}{rgb}{1,0,0}\definecolor{BLUE}{rgb}{0,0,1} %

\titleformat{\section}
{\normalfont\bfseries\centering}
{\Roman{section}.}{0.1in}{}
\titlespacing{\section}{0in}{0.25in}{0.1in}
\usepackage[labelfont=bf]{caption}

\usepackage{fancyhdr}
\fancyheadoffset{0pt}
\usepackage[%
style=phys,%
articletitle=false,biblabel=brackets,%
chaptertitle=false,pageranges=false%
]
{biblatex}
\title{ \textbf{Oscillating photonic Bell state from a semiconductor quantum dot for quantum key distribution}}
\date{}

\author[1]{ Matteo Pennacchietti}
\author[2]{ Brady Cunard}
\author[2]{ Shlok Nahar}
\author[3]{ Mohd Zeeshan}
\author[2]{ Sayan Gangopadhyay}
\author[3]{ Philip J. Poole}
\author[3,4]{ Dan Dalacu}
\author[5]{ Andreas Fognini}
\author[6]{ Klaus D. Jöns}
\author[7]{ Val Zwiller}
\author[2]{ Thomas Jennewein}
\author[2]{ Norbert Lütkenhaus}
\author[1,2]{ Michael E. Reimer}

\affil[1]{Institute for Quantum Computing and Department of Electrical and Computer Engineering, University of Waterloo, Waterloo, Ontario N2L 3G1, Canada}
\affil[2]{Institute for Quantum Computing and Department of Physics and Astronomy, University of Waterloo, Waterloo, Ontario N2L 3G1, Canada}
\affil[3]{National Research Council of Canada, Ottawa, Ontario K1A 0R6, Canada}
\affil[4]{University of Ottawa, Ottawa, Ontario, Canada K1N 6N5}
\affil[5]{Single Quantum B.V., Delft 2628 CJ, The Netherlands}
\affil[6]{Institute for Photonic Quantum Systems (PhoQS), Center for Optoelectronics and Photonics Paderborn (CeOPP) and Department of Physics, Paderborn University, 33098 Paderborn, Germany}
\affil[7]{Department of Applied Physics, Royal Institute of Technology, Albanova University Centre, 106 91, Stockholm, Sweden}

\renewenvironment{abstract}
 {
  \list{}{%
    \setlength{\leftmargin}{0.75in}
    \setlength{\rightmargin}{\leftmargin}%
  }%
  \item\relax}
 {\endlist}

\begin{document}
\twocolumn[
  \begin{@twocolumnfalse}
    \maketitle
    \vspace{-1cm}
    \begin{abstract}
    
    An on-demand source of bright entangled photon pairs is desirable for quantum key distribution (QKD) and quantum repeaters. The leading candidate to generate entangled photon pairs is based on spontaneous parametric down-conversion (SPDC) in a non-linear crystal. However, there exists a fundamental trade-off between entanglement fidelity and efficiency in SPDC sources due to multiphoton emission at high brightness, which limits the pair extraction efficiency to 0.1\% when operating at near-unity fidelity. 
    Quantum dots in photonic nanostructures can in principle overcome this trade-off; however, the quantum dots that have achieved entanglement fidelities on par with SPDC sources (99\%) have poor pair extraction efficiencies of 0.01\%. 
    Here, we demonstrate a 65-fold increase in the pair extraction efficiency compared to quantum dots with equivalent peak fidelity from an InAsP quantum dot in a photonic nanowire waveguide. 
    We measure a raw peak concurrence and fidelity of 95.3\% $\pm$ 0.5\% and 97.5\% $\pm$ 0.8\%, respectively. 
    Finally, we show that an oscillating two-photon Bell state generated by a semiconductor quantum dot can be utilized to establish a secure key for QKD, alleviating the need to remove the quantum dot energy splitting of the intermediate exciton states in the biexciton-exciton cascade.

    \end{abstract}
    \vspace{0.25in}
  \end{@twocolumnfalse}
]
\thispagestyle{empty}

Developing a bright, deterministic source of entangled photon pairs for applications in photonic quantum information processing \cite{Walther2005-in}, quantum communication \cite{Ekert1991-hk} and networks \cite{Rota2020-wc}, and enhanced imaging techniques \cite{doi:10.1126/science.1138007, PhysRevLett.118.257402, Lemos2014-dc} has been a long-standing scientific and technological challenge. 
One promising candidate is based on III-V semiconductor quantum dots (QDs) in photonic nanostructures \cite{Jons2017-oh, Liu2019-dz, Chen2018-nv, Wang2019-qy}. 
Such QDs have shown deterministic polarization-entangled photon pair emission via the biexciton (XX)-exciton (X) cascade and high pair-extraction efficiencies over 30\% \cite{Chen2018-nv, Liu2019-dz, Wang2019-qy}.
However, since semiconductor QDs are hosted in a solid-state environment, dephasing of the exciton spin was a common concern due to interactions with the surrounding nuclear spins of the atomic lattice and spins of free/trapped charges \cite{Kuhlmann2013-yq, Kuroda2013-rc}.
Therefore, it was presumed that because of indium's large nuclear spin of 9/2, indium-based QDs were unlikely to match the near-unity degree of entanglement (i.e., concurrence $\ge$ 95\%) of GaAs QDs, whose nuclei both have smaller nuclear spins of 3/2 \cite{Huber2017-px, Keil2017-nm}.

\begin{figure*}[t]
    \centering
    \includegraphics[width = 0.7\textwidth]{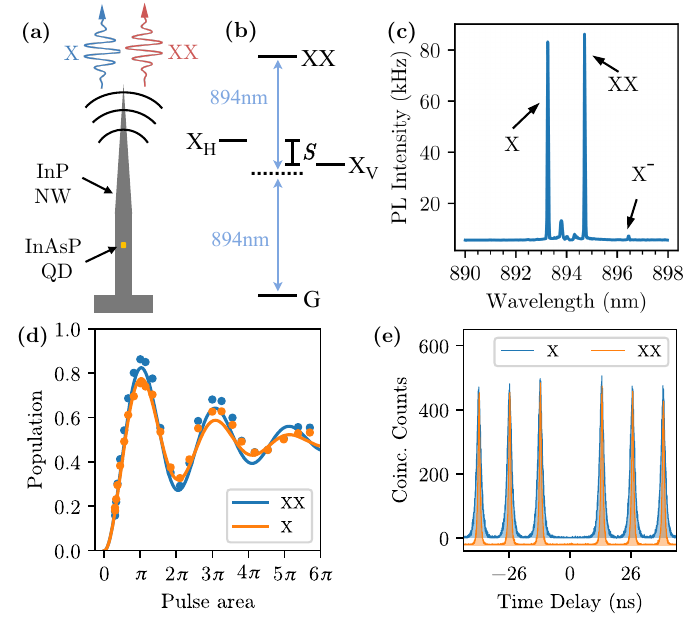}
   \vspace{-0.5cm}
    \caption{\textbf{Resonant excitation of the nanowire quantum dot.} (a) Illustration of the InAsP QD in an InP tapered nanowire waveguide, which emits the biexciton (XX) and exciton (X) into the same Gaussian mode. (b) Illustration of TPRE generating a biexciton in the QD. The intermediate exciton energy level has a finite fine structure splitting (S) between the H-polarized ($X_H$) and the V-polarized ($X_V$) states. (c) The emission spectrum from the nanowire QD under $\pi$-pulse TPRE. Exciton and biexciton emission lines are located at 893.25\,nm and 894.67\,nm, respectively. The small central peaks are from remnants of the filtered laser pulse midway between X and XX. We observe a small contribution from the negatively charged exciton (X$^-$). (d) Biexciton and exciton state population as a function of pulse area where a pulse area of $\pi$ corresponds to the maximum population. The dots represent experimental data and the solid curves are the fitted curves (see Supplementary Note \ref{supp_sec:rabi}). (e) Hanbury-Brown-Twiss autocorrelation measurements of both the X and XX emission, demonstrating low multi-photon emission probability.}
    \label{fig:figure1}
\end{figure*}

In contrast, it was suggested by Ref.~\cite{Fognini2019-uj} that indium-based QDs could reach near-unity degrees of entanglement by reducing the multi-photon emission probability, the single-photon detector dark count rate, and the detection system timing jitter. 
The latter metric is critical when the QD exciton fine-structure splitting (FSS) is non-zero, which causes the two-photon entangled state to oscillate between two Bell states.
Since previous studies did not combine an optimized detection system with two-photon resonant excitation (TPRE) to reduce the multiphoton emission probability \cite{Fognini2019-uj}, a concurrence of 95\% or greater has not yet been achieved with an indium-based QD.

In this article, we show near-unity degrees of entanglement from an InAsP QD in a site-selected tapered InP nanowire waveguide (NW-QD), see Figure \ref{fig:figure1}a. 
We measure a raw peak concurrence and fidelity of 95.3 $\pm$ 0.5\% and 97.5 $\pm$ 0.8\%, respectively.
This near-unity concurrence was achieved by performing a `time-resolved' quantum state tomography experiment (QST) using TPRE (Figure \ref{fig:figure1}b) and superconducting nanowire single-photon detectors (SNSPDs) with ultra-low timing jitter ($<$\,20\,ps) and dark count rate ($\sim$ 1\,Hz).
The time-resolved nature of the experiment enabled us to measure high values of concurrence over the entire exciton lifetime, resulting in a lifetime-weighted concurrence and fidelity of $90.2 \pm 0.2$\,\% and 94.0 $\pm$ 0.1\,\%, respectively.
These results correspond to a 27\,\% increase in the peak concurrence and a 48\% increase in the lifetime-averaged concurrence from previous measurements on the same NW-QD entangled photon source without using TPRE  \cite{Fognini2019-uj}. 
The stark increase validates the prediction that optimized experimental conditions will increase the measured concurrence on the same QD, and substantiates the initial claim that spin dephasing was not the factor limiting the measured entanglement fidelity in previous studies of NW-QD entangled photon sources \cite{Fognini2019-uj, Versteegh2014-sd, Huber2014-ii, Jons2017-oh}.
Furthermore, we propose a novel ``time-resolved'' quantum key distribution (QKD) protocol that is agnostic to the oscillations of the polarization entanglement caused by the energy splitting of the intermediate exciton states in the biexciton-exciton cascade. 
We further rigorously prove the security of our proposed time-resolved QKD protocol with more detailed optical models as compared to previous QKD experiments with QDs\,\cite{Schimpf2021-uq, Basso_Basset2021-rc,dzurnak2015quantum}.

Remarkably, our measurements show that these NW-QDs can outperform other indium-based QDs, and even those with no indium content \cite{Huwer2017-sx, Zeuner2021-lg, Hopfmann2021-ni}. 
Indeed, our raw peak measured concurrence is the highest in the literature to date.
Additionally, the pair-extraction efficiency of the source is over an order of magnitude higher than existing III-V QD sources with comparable entanglement fidelity \cite{Huber2017-px, Huber2018-dw}. %
This is an important finding because these NW-QD sources offer a number of attractive features, including deterministic site-selected growth \cite{Dalacu2012-ag}, intrinsically low exciton FSS \cite{Singh2009-lk}, potential pair-extraction efficiencies of over 90\% \cite{Claudon2013-wn, Reimer2012-mi}, near-unity fabrication yields of bright QDs \cite{Laferriere2022-aa}, Gaussian emission profile with $93\%$ coupling efficiency of single photons to a single-mode fibre \cite{Bulgarini2014-ma}, and wavelength emission variance from QDs of less than 5\,nm \cite{Chen2016-eu}.
These facts together with the high entanglement fidelity presented in this work illuminate a clear path for the deployment of large arrays of bright, entangled photon pair sources for quantum photonic technologies.

\section*{RESULTS}
\textbf{Two Photon Resonant Excitation}. 
Experimental results of TPRE on the NW-QD using laser pulses of $\sim$13\,ps (see Methods) are shown in Figure \ref{fig:figure1}.
In Figure \ref{fig:figure1}c, we observe that the exciton (X) and biexciton (XX) dominate the emission spectrum with the small peaks between X and XX corresponding to remnants of the laser excitation pulse. 
We observe three full Rabi cycles in the count rates vs. the pulse area in Figure \ref{fig:figure1}d, confirming that we are coherently exciting the biexciton state. 
The oscillations dampen out and settle to the 0.5 population point due to exciton-phonon interactions \cite{Luker2019-lc}.

We fit the count rate vs. pulse area data to determine an estimate for the probability of occupying the XX state after excitation (see Methods). 
At the $\pi$-pulse, we find a XX and X population of 0.82 $\pm$ 0.01 and 0.77 $\pm$ 0.01, respectively, resulting in a pair generation efficiency of 0.63 $\pm$ 0.01\%. 
Using the measured count rates of 150\,kHz for XX and 145\,kHz for X under pulsed excitation (76.2\,MHz), and the optical setup efficiency of 2.4\%, we estimate a pair extraction efficiency at the first lens of 0.65\% $\pm$ 0.02\% (see Methods). 
The pair extraction efficiency is a 65-fold increase from a GaAs QD with comparable peak concurrence \cite{Trotta_ent_fid}.

To quantify the multi-photon emission probability under TPRE we perform a Hanbury-Brown-Twiss experiment. 
The resulting histogram is displayed in Figure \ref{fig:figure1}e. 
Using only the nearest neighbour peaks we calculate $g^{(2)}_{X}(0) = 0.0055 \pm 0.0003$ and $g^{(2)}_{XX}(0) = 0.0028 \pm 0.0003$.
We observe that $g^{(2)}_{XX}(0)$ is reduced by over an order of magnitude by TPRE as compared to previous work on the same NW-QD using quasi-resonant excitation ($g^{(2)}_{XX}(0) = 0.10 \pm 0.01$) \cite{Fognini2019-uj}.

\textbf{Influence of Detection System}.
In addition to the excitation technique, there are two main performance metrics of the detector that degrades the measured degree of entanglement. 
The first is detector dark counts, which is well understood and adds uncorrelated coincidences similar to multi-photon emission.
The second is the finite timing resolution of the detection system (i.e., timing jitter), which is largely unexplored and can reduce the measured concurrence when the exciton FSS is non-zero. 
The FSS is a splitting of the two intermediate exciton states \cite{Huber2018-dw} that causes the exciton spin state to oscillate between $\ket{\uparrow\Downarrow}$ (-1 angular momentum) and $\ket{\downarrow\Uparrow}$ (+1 angular momentum) with a period of $T_S = h/S$, where $S$ is the energy of the FSS.
The resulting polarization state of the photon pair emitted via the biexciton-exciton cascade will evolve temporally as:
\begin{equation}
\label{eq:bellstate-fss}
\begin{split}
        \ket{\psi(\tau)} & = \frac{1}{\sqrt{2}}\left[ \cos\big( S\tau/2\hbar \big) \Big(\ket{RL} + \ket{LR}\Big) \right. \\
        & \left. +i \sin\big( S\tau/2\hbar \big)\Big(\ket{RR} + \ket{LL}\Big) \right] \\
        & = \frac{1}{\sqrt{2}}\left(\ket{HH} + e^{i S \tau / \hbar }\ket{VV}\right) ,
\end{split}
\end{equation}
where $\tau = t_{X} - t_{XX}$ is the time delay between the emission of the exciton photon and the biexciton photon.
The first and second letter denotes the polarization of the biexciton and exciton photon, respectively.
The effect of the FSS is a purely local unitary transformation of the exciton spin, implying the state in Eq. (\ref{eq:bellstate-fss}) is maximally entangled over the entire exciton lifetime \cite{Fognini2019-uj}. 
However, if the finite timing resolution of the detection system is larger than the FSS oscillation period, then the measured concurrence will be reduced since the phase relationship between $\ket{HH}$ and $\ket{VV}$ in Eq. \ref{eq:bellstate-fss} will be averaged out. This effect can be circumvented by using single-photon detectors with a timing resolution much less than the FSS oscillation period.

\begin{figure}[t]
    \centering
    \includegraphics[width = \columnwidth]{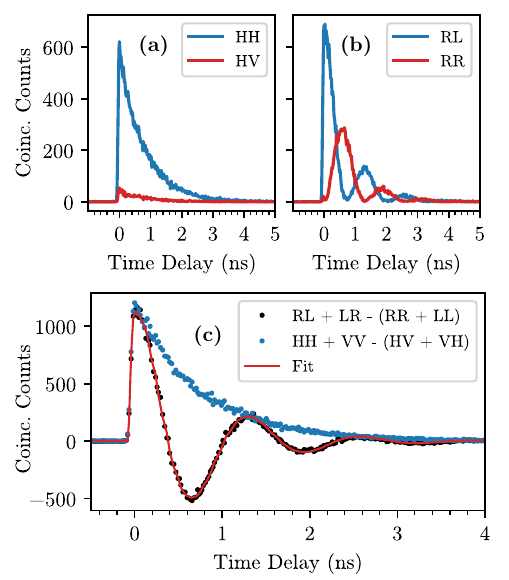}
    \caption{\textbf{Time-resolved quantum state tomography with SNSPDs.} 
    (a) Coincidence counts versus time delay ($\tau$) for the HH and HV polarization projection measurements.
    (b) Coincidence counts versus time delay ($\tau$) for the RL and RR polarization projection measurements. 
    (c) Combinations of coincidence measurements for the states in Eq. \ref{eq:bellstate-fss} versus time delay. 
    The combinations of circular basis measurements $RL + LR - (RR + LL)$ display quantum oscillations between two Bell states.
    The red curve is the best fit of the oscillations in the circular basis to $\cos{\left( S \tau /\hbar\right)}$ convolved with a Gaussian distribution of fixed FWHM of 30\,ps to account for the small amount of blurring from the SNSPD detection system.}
    \label{fig:figure2}
\end{figure}

\textbf{Quantum State Tomography}. 
We carried out an over-complete quantum state tomography (QST) \cite{James2001-vw} experiment using TPRE and fast SNSPDs with individual detector timing jitters of less than 20\,ps and very low dark counts rates of $N_d = 1$ $\pm 1$\,Hz.
Taking into both SNSPDs and the correlation electronics, we estimated a total system timing jitter (FWHM) of 30\,ps (see Methods), which is much faster than the exciton FSS period of the QD under study ($T_S = 1.26$\,ns).
The QST experiment consisted of a total of 36 separate projective measurements on the joint biexciton-exciton polarization state (see Supplementary Figure \ref{fig:SNSPD_individual} and \ref{fig:APD_individual}).
A subset of the 36 projection measurements captured using the SNSPDs is plotted in Figure \ref{fig:figure2}a (HH, HV) and Figure \ref{fig:figure2}b (RL, RR).
The coincidence counts from HH follow an exponential decay since the exciton spin state $\ket{\uparrow\Downarrow}$ + $\ket{\downarrow\Uparrow}$ is a stationary state of the exchange Hamiltonian when the FSS is present \cite{Bayer2002-xo}.
We see high suppression of coincidences in the HV basis as compared to the HH basis, as expected for a highly entangled state.
In contrast, the RL and RR states are not eigenstates of the exchange Hamiltonian when the FSS is present. 
Thus, the coincidences histogram for the RR and RL basis, displayed in Figure \ref{fig:figure2}b oscillates $\pi/2$ out of phase with each other.

Eq. \ref{eq:bellstate-fss} predicts the oscillation between two Bell states, $(\ket{RL}+\ket{LR})/\sqrt{2}$ and $(\ket{RR}+\ket{LL})/\sqrt{2}$, with frequency $S/h$.
We found that plotting the individual measured coincidences in the form of $RL\,+LR\,-\,(RR\,+\,LL)$ reveals ``quantum oscillations'' between these two Bell states (Figure \ref{fig:figure2}c).
We extract the FSS by fitting the quantum oscillations (red curve in Figure \ref{fig:figure2}c), yielding $3.226 \pm 0.004$ $\mu$eV ($780.0 \pm 1.0$ MHz). %
In contrast, the combination of $HH + VV - (HV + VH)$ follows an exponential decay. 
By fitting this decay, we extract the radiative exciton lifetime of $\tau_X = 0.777 \pm 0.003$ns (see Supplementary Figure \ref{fig:lifetime_fit}).

\begin{figure}[b!]
    \centering
    \includegraphics[width = 1\columnwidth]{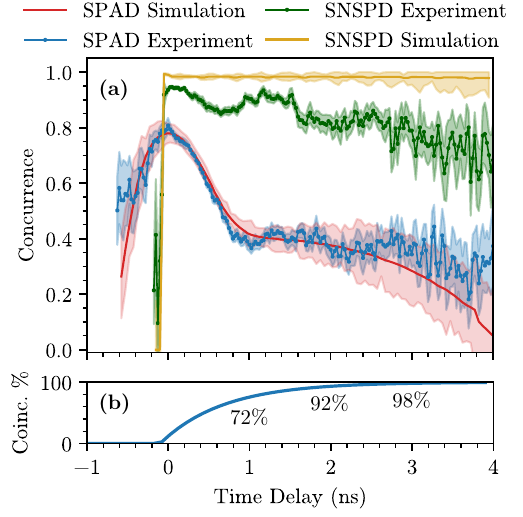}
    \caption{\textbf{Near-unity concurrence from a nanowire quantum dot.} (a) Concurrence from reconstructed density matrices as a function of time delay for experimental data and simulated data with time windows of $50$ps. Data is presented for NW-QDs excited under TPRE at $\pi$-pulse using both SPADs and SNSPDs. The shaded regions represent one standard deviation. (b) Percentage of total coincidence counts as a function of time delay.}
    \label{fig:figure3}
\end{figure}
\vspace{1cm}
\textbf{Concurrence under Resonant Excitation}. 
With the 36 time-resolved coincidence measurements we reconstruct the ``instantaneous'' density matrix, within a time window of 50\,ps as a function $\tau$, using an algorithm based on the maximum likelihood estimation (MLE) \cite{James2001-vw} implemented in Python \cite{mle-python}.
From these reconstructed density matrices we calculate the time-resolved concurrence \cite{Wootters2001-vr}.
In Figure \ref{fig:figure3}a we plot the concurrence versus time delay for measurements taken using TPRE and SNSPDs with low dark counts and precise timing jitter (green curve).
As a comparison, we also plot the measured concurrence (blue curve) on the same QD using TPRE and single-photon avalanche diodes (SPADs), which have significantly higher dark count rates and timing jitter (see Methods). %
The prominent effect of the detection system on the measured concurrence on the same QD is immediately evident from the difference in the two measured curves acquired using SNSPDs versus SPADs. 

Due to the large timing jitter of the SPADs (see Supplementary Figure \ref{fig:DetectionResponseFunction}), XX-X coincidences are recorded when $\tau < 0$. 
As $\tau$ increases and more XX-X coincidences are sampled, there is an initial slow rise of the SPAD concurrence.
The increase then slows and reaches a peak of $ 80 \pm 3\% $ due to phase averaging or ``blurring'' of the quantum oscillations, i.e., time-dependent phase information between the two Bell states in Eq. \ref{eq:bellstate-fss}. 
The concurrence then steadily decreases as the detection system samples more of the quantum oscillations, eventually reaching the ``flat'' region where the entire detection response function has sampled the temporal evolution of the two-photon entangled state.
In stark contrast, there is no distinction between the ``top'' and ``flat'' regions in the SNSPD concurrence curve.
Instead, at $\tau = 0$ the concurrence (fidelity) peaks at $\mathcal{C}_p = 95.3 \pm 0.5\,\%$ ($F_p = 97.5 \pm 0.8\,\%$), an increase of 24\% in the concurrence from the SPAD experiment.
We found that taking into account our measured multi-photon emission probability has little effect on the peak concurrence (95.6 $\pm$ 0.7\%) and fidelity (97.7 $\pm$ 0.4\%).
For the time-resolved entanglement fidelity, see Supplementary Figure \ref{fig:fid_osc_dm_plots}.
The concurrence then remains relatively flat for $\tau > 0$.
As a result, the lifetime weighted concurrence (fidelity) with the SNSPDs was $\Tilde{\mathcal{C}} = 90.2 \pm 0.2\,\%$ ($\Tilde{F} = 94.0 \pm 0.1\,\%$), which is significantly higher than the lifetime weighted concurrence of $50.8 \pm 0.2\,\%$ acquired with the SPADs.

After $\tau \ge 2$\,ns we enter the ``roll-off'' region where the concurrence reduces because the signal-to-dark count ratio begins to decrease as dark counts dominate the coincidences. Again, we notice a distinct difference between the SPAD and SNSPD curves, where the former decreases much more than the latter because the dark count rate of the SPADs (tens to hundreds per second) is significantly higher than the dark count rate of the SNSPDs (one per second).

Additionally, in Figure \ref{fig:figure3}a we plot the time-resolved concurrence versus time delay calculated from a ``dephasing-free'' model for the SPADs (red) and SNSPDs (yellow). 
The dephasing-free model accounts for only experimentally measured parameters: the multi-photon emission probability, the measured count rates of X and XX, the detector dark count rate, and the detector system timing response (see Supplementary Note \ref{supp_sec:dephasing_model}).
We see excellent agreement between the model and the experimental concurrence for the SPADs, with the predicted peak concurrence of $78 \pm 3$\,\%. 
In contrast, the simulation for the SNSPDs differs from the experimental data in three areas: (1) at $\tau = 0$\,ns; (2) $0 <\tau < 2$\,ns; and (3) $\tau > 2$\,ns.

(1) At $\tau = 0$\,ns. The small discrepancy between the experimental peak concurrence and the $99.1 \pm 0.04 \, \%$ concurrence predicted by the model could be due to the combination of leakage excitation laser photons (see Supplementary Figure \ref{fig:polarized_hbt}), the interaction of the exciton spin with phonons \cite{Hohenester2007-eb, Cygorek2018-ev} and/or the large energy splitting caused by resonant excitation via the AC Stark Shift \cite{Seidelmann2022-my}.
The first mechanism would result in unwanted correlations from the detection of H-polarized laser photons.  
The latter two mechanisms act on time scales less than the timing resolution of our detection system, thereby resulting in a small reduction in the measured concurrence.

(2) $0 <\tau < 2$\,ns. We note that there is a small oscillation in the measured concurrence between $\tau = 0$ and $\tau = 1.3$\,ns, which coincides with the FSS period.
This drop in concurrence is not from spin-dephasing since it fully recovers around 1.3\,ns.
Instead, the concurrence dip might be attributed to the QD emission being slightly polarized due to asymmetry in the nanowire shape and/or the waveplates not being perfect quarter and half.

(3) $\tau > 2$\,ns. The divergence of the experimental data from the model in the ``roll-off'' region can be explained by spin-scattering of the exciton spin before recombination with a characteristic interaction time of $\sim 25$\, ns (Supplementary Figure \ref{fig:dephasing}). 
Such a dephasing time is over an order of magnitude greater than the exciton lifetime, which is in agreement with previous experimental evidence from the neutral exciton in InGaAs QDs \cite{Senes2003-pn, Kuhlmann2013-yq}. 
We note that with the SNSPDs we were able to observe spin-dephasing from InAsP QDs in nanowires, which was not possible with SPADs due to their higher timing jitter.
Additionally, as illustrated in Figure \ref{fig:figure3}b, only 8\% of the exciton photons are emitted in this region where the concurrence drops.

\textbf{Time-resolved Entanglement-based QKD}.
Given that we observe high entanglement fidelity over the entire exciton lifetime, we ask: Is it possible to perform a quantum information task using the oscillating quantum state from Eq. \ref{eq:bellstate-fss} without temporal post-selection?
Of course, if the quantum information protocol of interest requires one specific Bell state, the oscillation caused by the FSS is an obstacle.
However, for quantum key distribution (QKD) a specific Bell state is \textit{not} required.

Indeed, we can devise a ``time-resolved" QKD protocol where access to a maximally entangled state of the form $\left(\ket{HH}+\exp(i S\tau/\hbar) \ket{VV}\right)/\sqrt{2}$ is sufficient to generate a secret key. 
In this case, we would map $H$ polarized light to the bit $0$, and $V$ to the bit $1$.
This key map from quantum to classical data is expectedly agnostic to the phase of the entangled state.
As detailed in Supplementary Note \ref{supp:QKD}, in addition to the key map, the procedure to estimate the state shared between Alice and Bob in the QKD protocol is similar to quantum state tomography.
Therefore, since a detection system with low-timing jitter compared to the FSS oscillation period allows us to observe a high concurrence through tomographic reconstruction, we prove in Supplementary Note \ref{supp:QKD} the state at every $\tau$ is able to generate a ``time-resolved key rate''. We further show that our time-resolved analysis is valid for any time-dependent states.

\begin{figure}[t]
    \centering
    \includegraphics[width = \columnwidth]{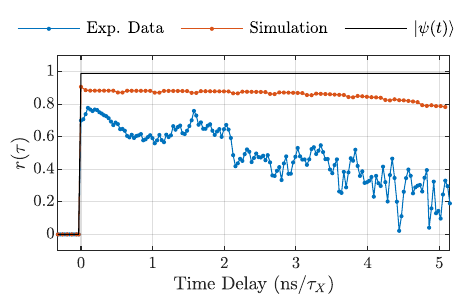}
    \caption{ \textbf{Time-resolved key rate.}  Theoretical key rate for: experimental data (blue) with optimal choice of measurement bases states, dephasing-free model (red) and ``perfectly oscillating'' state given by Eq. \ref{eq:bellstate-fss} (black).}
    \label{fig:qkd_keyrate}
\end{figure}

As an example, we consider a time-resolved optical implementation of the 6-state protocol \cite{bruss1998optimal}. 
We compute the time-resolved key rate for this protocol, i.e., the number of secret bits produced per two-fold coincidence as a function of time delay $\tau$.
This is visualized in Figure \ref{fig:qkd_keyrate}, where we compare the key rates for the raw data from the NW-QD (blue curve) with the ideal state given in Eq. \ref{eq:bellstate-fss} (black curve) and the SNSPD simulation from Figure \ref{fig:figure3}a (red curve).
From Figure \ref{fig:qkd_keyrate}, we see that the time-resolved key rate is positive up to time delays between the biexciton and exciton emission of $5 \tau_X$\,ns, implying that all photon pairs from the NW-QD are useful for key generation.

The deviation of the simulated and experimental time-resolved key rate from the perfect state ($r(\tau) = 0.99$) shows that the time-resolved key rate is a more sensitive metric than concurrence since $r(\tau)$ is dependent on the closeness to maximally entangled states in a subspace spanned by Alice and Bob's key map choices. 
For example, the black curve in Figure \ref{fig:qkd_keyrate} for the perfect state gives the maximum possible key rate as the state oscillates perfectly in the plane spanned by $HH$ and $VV$. 
However, the oscillations in the blue curve imply there is no such choice of key map (see Supplementary Notes \ref{supp-sec:time-resolved_ent_fidelity} and \ref{supp:QKD}) for the experimental data.
Similarly, the non-zero multi-photon emission probability and finite dark count rate in the simulation, causes the red curve to sit below the perfect state.  

We can calculate the number of bits of the secret key produced per excitation pulse $R$ by calculating a lifetime-weighted time-resolved key rate.
When averaging from $\tau = 0$\,ns to $\tau = 5\tau_X$\,ns we find $R = 0.88$ and $R = 0.64$ for the dephasing-free model and the experimental data, respectively. Note that we have assumed a perfect channel for our proof-of-principle analysis, so the key rate exactly reflects the utility of the source for QKD. In practice, the channel would add some errors and loss which would further degrade this key rate.

\section*{DISCUSSION}

It is well understood that the exciton FSS will reduce the measured entanglement fidelity defined with respect to one specific Bell state unless temporal post-selection is used  \cite{Kuroda2013-rc, Chen2018-nv, Keil2017-nm, Huber2018-as}.
Thus, if the FSS is not directly removed, then one must settle with either a less efficient entangled photon source or a lower measured entanglement fidelity.
In both cases, the overall performance of the entangled photon source is reduced.

Our QKD analysis shows that this trade-off between efficiency and fidelity can be overcome by removing the assumed constraint of requiring only one specific Bell state.
Indeed, we constructed a time-resolved QKD protocol where \textit{all} photon pairs emitted by a QD with non-zero FSS can be used in secret key generation.
This protocol works only when the detection system's temporal resolution is much smaller than the FSS period. 
By implementing our protocol, the key rates in previous QKD experiments with QD entangled photon pair sources \cite{Schimpf2021-uq, Basso_Basset2021-rc, dzurnak2015quantum} could be improved since these experiments did not account for the FSS.
Additionally, unlike previous security analyses that assume perfect qubit states \cite{Schimpf2021-uq, Basso_Basset2021-rc, dzurnak2015quantum}, we rigorously bound the effect of any multi-photon components of the optical state on the key rate, which is more applicable to practical implementations.

The core experimental challenge with this time-resolved QKD scheme is ensuring accurate timing synchronization on the order of 10-20\,ps between the two parties.
Such accurate timing synchronization is within reach of current technology since it was recently shown that two distant quantum-networked nodes were synchronized to $<$ 10\,ps over a few kilometres \cite{Gerrits2022Conf.LasersElectro-Opt.}.
Thus, it may be less challenging to satisfy the synchronization requirements instead of directly removing the FSS which requires complex fabrication methods \cite{Huber2018-as, Zeeshan2019-na} or external optics \cite{Fognini2018-sp}.
Therefore, our time-resolved protocol could significantly relax the technical requirements for using bright QD entangled photon sources for QKD.

We found that our QD entangled photon source suffers from significant luminescence blinking (see Supplementary Figure \ref{fig:blinking}) due to background n-doping of the nanowire (see Supplementary Information of Ref. \cite{Reimer2016-zm}).
It has been shown in previous work that blinking can be completely removed via the application of an electric field to sweep away excess carriers from the vicinity of the QD \cite{Zhai2020-yh}.
If blinking is successfully removed, our pair extraction efficiency would increase almost six-fold (see Supplementary Note \ref{supp_sec:eff_est}).
Additionally, techniques to improve the pair generation rate from 80\,\% to unity can be achieved by optimizing the TPRE pulse \cite{Kaldewey2017-ia}.
Finally, the addition of a gold mirror at the nanowire base should lead to a further two-fold boost of the pair extraction efficiency \cite{Reimer2012-mi}.

In summary, we have shown that NW-QD emitters are a bright source of entangled photon pairs with concurrence values that are on par with other leading semiconductor QD devices \cite{Zeuner2021-lg, Hopfmann2021-ni, Huwer2017-sx, Huber2018-dw, Liu2019-dz, Wang2019-qy}. 
Our results demonstrated that the low concurrence values found in earlier studies of NW-QD emitters were not due to spin dephasing from the large 9/2 nuclear spin of indium.
Instead, imperfect experimental conditions, namely multi-photon emission, detector dark counts, and timing jitter, were the limiting factors.
Additionally, we found that any residual spin dephasing occurred over time scales that were much longer than the exciton's radiative lifetime.
Our findings lay the foundation for creating large arrays of deterministically-positioned NW-QDs for QKD networks.

\section*{METHODS}
\subsubsection*{Quantum Dot Source}

Please refer to Ref. \cite{Versteegh2014-sd} for a description of the nanowire quantum dot growth procedure.

\subsubsection*{Tomography Apparatus}

A standard cryogenic micro-photoluminescence (PL) apparatus was used to maintain the nanowire sample at 4.5K and enable the optical interfacing via a top window and mirror.
Excitation pulses were directed to a 70:30 beamsplitter, with the reflection passing to the micro-PL cryostat.
The excitation pulse power was monitored with a photodiode on the transmission port of the 70:30 beamsplitter. 
After excitation, the photons from exciton and biexciton transition, which are separated by 1\,nm in wavelength, coupled to the same waveguide mode of the nanowire and then propagate out into free space. 
The emission from the nanowire is collimated with the objective lens and directed up out of the cryostat.
The emission travels through the 70:30 beamsplitter and then three notch filters which filter out the light from the resonant pulse. Then, a transmission grating is used to separate the exciton and biexciton photons into different spatial modes. 
With the photons separated, we implemented a quantum state tomography setup similar to \cite{Huber2014-ii}.
Independent polarization transformations were performed on each photon using a $\lambda$/4 followed by a $\lambda$/2 waveplate.
Projective polarization measurements were then performed using a polarizer.
All four waveplates were mounted inside high-precision, stepper-motorized rotating stages which allowed for automated data-taking and the precise setting of the waveplate principle axes to implement each measurement. 
The beam was then coupled to a single model fibre and passed through a fibre-based fibre-based bandpass filter (FWHM of 0.07\,nm) to the single-photon detectors.
The electrical signals produced by the detection events are then sent to the 2-channel time-tagging electronics. The histogram bin width was set to 16\,ps and the integration time for capturing each histogram was 300 seconds. Please see Supplementary Figure \ref{supp_sec:exp_apparatus} for an illustration of the experimental apparatus. 
Please see Supplementary Note \ref{supp_sec:apparatus_calibration} for information on how we calibrated the waveplates and determined the measurement settings.  

\subsubsection*{TPRE Pulse}

For TPRE, a 3\,ps pulse from a Mira 900P Ti:Sapphire laser was extended to an approximately 13\,ps pulse using a standard 4-f pulse shaper with two 1800 grooves/mm gratings. We note that the temporal duration was not measured directly, instead, we are assuming that the pulse is Fourier-limited, a reasonable assumption to make in the absence of any mechanism introducing chirp into the setup. We have measured a spectral linewidth (FWHM) of 0.08\,nm at 894\,nm, thus by assuming a Gaussian pulse envelope, the time-bandwidth product gives $\Delta t \approx 0.441/\Delta \nu$.

\subsubsection*{Detection System}
For the detection system, the two SNSPDs (Single Quantum Eos) had a timing resolution of 19\,ps and 18\,ps, respectively, with the correlation electronics (PicoQuant PicoHarp 300) having a timing resolution of 10\,ps.
Each exhibits a Gaussian distribution resulting in an overall timing resolution of $\sqrt{19^2 + 18^2 + 2*10^2} \approx 30$\,ps. 
The overall detection system timing resolution when using the APDs was 488 $\pm$ 1\,ps, see the Supplementary Note \ref{supp_sec:spad_timing_response} for the calculation of this value. The two APDs had dark count rates of 34 $\pm$ 18 and 306 $\pm$ 51, respectively.

\subsubsection*{Source Efficiency}
To calculate the efficiency under TPRE, we sent the emission of the nanowire sample directly to a high-resolution diffraction spectrometer.
A single SPAD detector was placed at the output of a controllable output split inside the spectrometer to control the selected wavelength of the detector.
The efficiency of the collection objective and top mirror of the cryostat is 80\% and 95\%, respectively. 
The transmission of the beamsplitter used for mixing in the excitation pulse is 65\%.
The total optical efficiency after the 70:30 beamsplitter (see Supplementary Figure \ref{fig:exp_apparatus}) to the SPAD detector after the spectrometer was estimated to be 2.41 $\pm$ 0.07\%. 
The unpolarized count rates for both the X and XX emission line were sent to the spectrometer, and the slit was used to select either to record the total count rates in each line.
We found 150k counts/sec for the biexciton and 145k counts/sec with the exciton.
This corresponds to a pair extraction efficiency at the first lens of 0.65 $\pm$ 0.02\%.

\subsubsection*{Lifetime-Averaged Metrics}
To calculate the lifetime average we use the equation,
\begin{equation}
    \Tilde{\mathcal{C}} = \frac{ \sum_{\tau_n} N_{\tau_n} \mathcal{C}(\rho(\tau_n))} {\sum_{\tau_n} N_{\tau_n} }
\end{equation}
where $\tau_n$ is the time bin of the coincidence histogram, $\mathcal{C}(\rho(\tau_n))$ is the concurrence of the density matrix $\rho$ reconstructed at time bin $\tau_n$ and $N_{\tau_n} = \sum_{ij} N_{\tau_n}^{ij}$ is the sum of all coincidences in the 36 polarization measurements $i,j \in \{H,V,D,A,R,L\}$.
Equivalently, we use a similar formula to calculate the key rate
\begin{equation}
    R = \frac{ \sum_{\tau_n} N_{\tau_n} r(\tau_n)} {\sum_{\tau_n} N_{\tau_n} }
\end{equation}
where $r(\tau_n)$ is the time-resolved key rate calculated for time delay $\tau_n$. The theoretical basis for this is expanded on in Supplementary \ref{supp:QKD}.

\subsection*{Acknowledgements}
This research was undertaken thanks in part to funding from the Canada First Research Excellence Fund, NSERC (programs Canadian Graduate Scholarship - Masters and Discovery), CMC Microsystems and the Mitacs Accelerate program.
We acknowledge the generosity of Single Quantum B.V. for providing us with the superconducting nanowire single-photon detection system for this work. 

\subsection*{Author contributions}
M.P. performed the experiments with significant contributions from B.C. and M.Z.
Data analysis and modelling were performed by M.P. with code developed by A.F. 
The development and fabrication the quantum dot emitter was done by D.D. and P.J.P.
The development of the QKD protocol was led by S.N. under the supervision of N.L. with assistance from S.G.
Two-photon resonant excitation of the quantum dot was implemented by K.D.J.
M.E.R. supervised the project and wrote the paper with M.P. and S.N. with input from all authors.

\subsection*{Competing Interests}
A.F. is employed by Single Quantum and may profit financially. V.Z. is the co-founder and chief scientific officer of Single Quantum and may profit financially. The other authors declare no competing interests.

\printbibliography

\begin{refsection}
\onecolumn
\counterwithin{figure}{section}
\refstepcounter{section}
\pagenumbering{arabic}
\renewcommand{\thepage}{S\arabic{page}} 
\renewcommand{\thesection}{S\arabic{section}}  
\renewcommand{\thetable}{S\arabic{table}}  
\renewcommand{\thefigure}{S\arabic{figure}}
\renewcommand{\theequation}{S\arabic{equation}}

\section*{SUPPLEMENTARY INFORMATION}

\subsection{Fitting Rabi Oscillations}\label{supp_sec:rabi}
To fit the Rabi oscillations to a population of the biexciton state we used the formula
\begin{equation}
\label{eq:rabi-fit-fox}
    |c_e|^2 = \frac{1}{2(1 + 2\xi^2)} \left( 1 - \left( \cos(\Omega' t) + \frac{3\xi}{\sqrt{4 - \xi^2}} \sin(\Omega' t) \right) e^{-3\gamma t/2} \right)
\end{equation}
from Ref. \cite{Fox}, where $\Omega' = \Omega_R \sqrt{1 - \xi^2/4}$ and $\xi = \gamma/\Omega_R$, with $\gamma$ representing a general damping rate parameter and $\Omega_R$ being the standard Rabi frequency.
The Rabi oscillation data in Figure \ref{fig:figure1}a) of the main text is recorded as a function of the average excitation pulse power at a fixed temporal pulse width.
Thus, the Eq. \ref{eq:rabi-fit-fox} above is not useful as it depends on time, a parameter we do not have direct access to. 
Recall, that in the regime of pulsed resonant excitation, we define the \textit{pulse area}
\begin{equation}
    \Theta = \abs{ \frac{\mu_{21}}{\hbar} \int \mathcal{E}_0(t) dt}
\end{equation} 
where $\mu_{21}$ representing the dipole moment of a given resonant transition and $\mathcal{E}_0(t)$ is the temporal envelope of the excitation pulse.
For a constant field $\mathcal{E}_0$, the pulse area is therefore exactly $\Omega_R t$ and 
thus, in Eq. \ref{eq:rabi-fit-fox} we can take $\Omega_R t \rightarrow \Theta$.
Therefore we can substitute $\Omega' t \rightarrow \Theta \sqrt{1 - \xi^2/4}$ and $\gamma t = \Omega_R \xi t \rightarrow \Theta \xi$ into Eq. \ref{eq:rabi-fit-fox} and find 
\begin{equation}
    |c_e|^2 = \frac{1}{2(1 + 2\xi^2)} \left( 1 - \left( \cos(\Theta) + \frac{3\xi}{\sqrt{4 - \xi^2}} \sin(\Theta) \right) e^{-3\Theta\xi/2} \right) .
\end{equation}
This equation depends only on the parameters $\Theta$ and $\xi$, and can therefore be used to fit our data directly.

\newpage

\subsection{Raw Projections}\label{supp:raw_projections}
\begin{figure}[H]
    \centering
    \includegraphics[width = \textwidth]{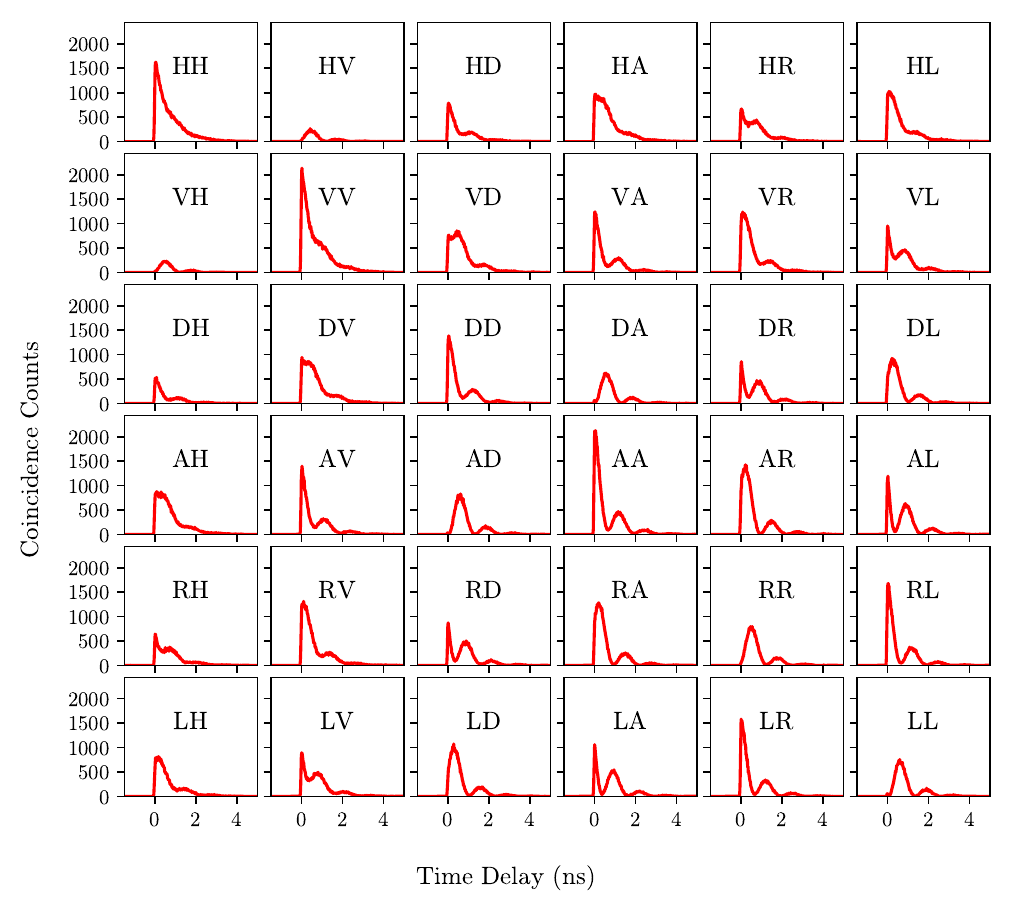}
    \caption{Raw coincidence histograms of each of the 36 projective measurements using the SNSPDs.}
    \label{fig:SNSPD_individual}
\end{figure}

\begin{figure}[t!]
    \centering
    \includegraphics[width = \textwidth]{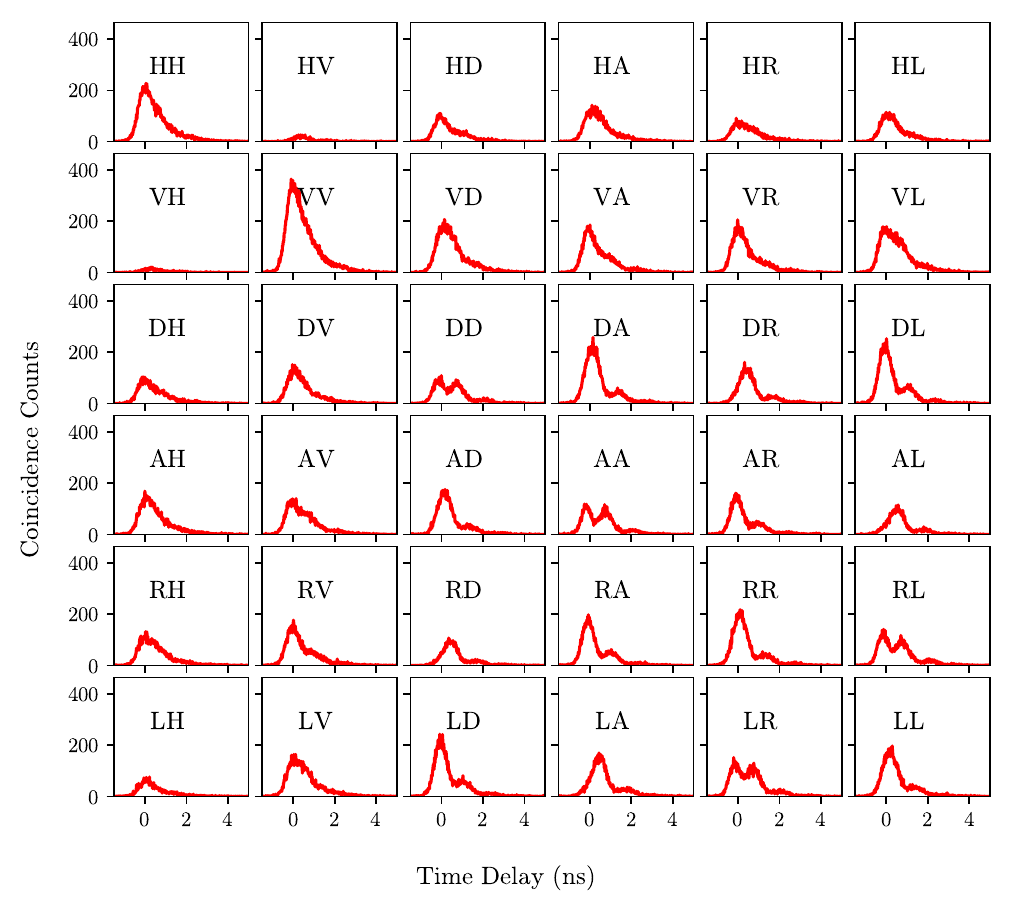}
    \caption{Raw coincidence histograms of each of the 36 projective measurements using the SPADs.}
    \label{fig:APD_individual}
\end{figure}

\clearpage

\subsection{Exciton Lifetime}
\begin{figure}[h!]
    \centering
    \includegraphics[width = 0.5\textwidth]{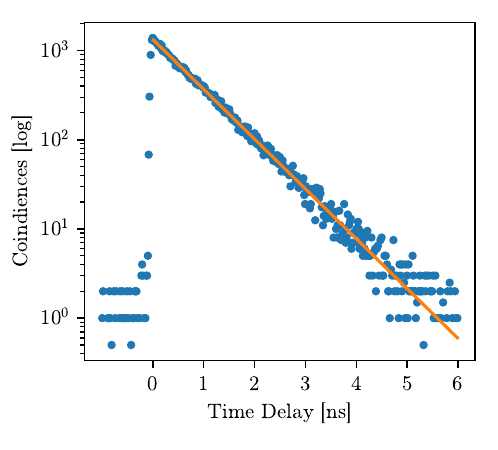}
    \caption{A plot of the raw biexciton-exciton cross-correlation coincides vs. time delay $t_{X} - t_{XX}$ with the coincidences on a log scale. Raw data is indicated by the blue dots and the exponential fit by the orange curve. }
    \label{fig:lifetime_fit}
\end{figure}

To determine the lifetime of the exciton state, we must look at biexciton-exciton cross-correlations vs. the detection time delay, i.e., the time between the detection of a biexciton and the subsequent detection of an exciton: $\tau = t_{X} - t_{XX}$.
The causal nature of the cascade, i.e., in the case of resonant an exciton can only form after the biexciton is emitted, means that the decay of coincidences over this time difference will correspond directly to the true radiative lifetime of the exciton state.
In Figure \ref{fig:lifetime_fit} we plot the sum of the coincidences corresponding to all four possible configurations that the cascade can emit into, i.e., $RR + LL + RL + LR$.
We fit these correlations to a simple exponential decay function $A\exp(-\tau/\tau_X)$ where $t$ is the time delay discussed and $\tau_X$ is the lifetime of the exciton.
As stated in the main text we find a value of $0.777 \pm 0.003$ ns.
The exponential decay model shows very close agreement with the data.

\subsection{SPAD Timing Response} \label{supp_sec:spad_timing_response}
\begin{figure}[h!]
    \centering
    \includegraphics[width = \textwidth]{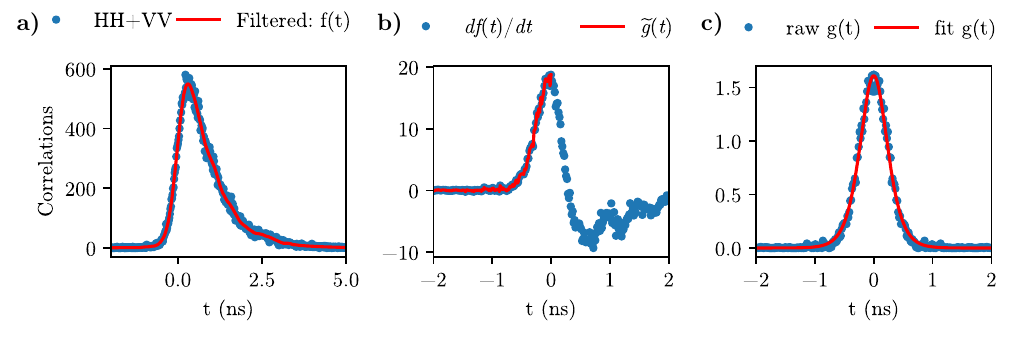}
    \caption{a) Raw $HH + VV$ projection coincidence counts (blue dots) with a Savitzki-Golay filter (red curve) with a polynomial degree of 6 and a window length of 65, applied in order to smooth the raw data. b) Numerical derivative of the filtered curve (blue dots) with only points where $t \leq 0$ plotted in red and denoted as $\Tilde{g}$. c) Mirrored version of $\Tilde{g}$ across the x-axis to generate the raw data that represents $g(t)$ (blue points), which is then fitted to Eq. \ref{eq:timing-response} (red curve). All plots are against the coincidence time delay, denoted as t.}
    \label{fig:DetectionResponseFunction}
\end{figure}

Our avalanche photodiodes (SPADs) are not the same as those employed in Ref. \cite{Fognini2019-uj}, but the procedure for extracting the timing resolution of the detection system is the same.
The results of the analysis done on our SPAD detection system are displayed in Figure \ref{fig:DetectionResponseFunction}.
The $HH$ and $VV$ biexciton-exciton cross-correlation coincidence histograms should follow a simple exponential decay function for $t > 0$ with a Heaviside step function $\Theta(t)$ to represent the causal nature of the cascade, i.e., the exciton is always created after the recombination of the biexciton. 
The histogram for the sum $HH$ + $VV$ histogram is then the convolution of the above function with the detection timing response $g(t)$
\begin{equation}
    f(t) = \Theta(t) e^{-t/\tau_X} * g(t) .
\end{equation}
The detection response function will thus be imprinted onto the coincidence histogram data at time delays $ t \leq 0$, i.e., the rising edge of the curve, where $\Theta(t)$ dictates the shape.
We can therefore limit ourselves to histogram data points where $ t \leq 0$,
\begin{equation}
    f(t \leq 0) \approx \Theta(t) * g(t) = f(t_-) \implies \frac{d f(t_-)}{dt} = d\left[ \Theta(t) * g(t) \right] = d\left[ \Theta(t) \right] * g(t) = \delta(t) * g(t) = g(t). \, \footnote{ Recall that the derivative of the convolution function is simply the convolution between the derivative of one of the functions, and the other function, i.e., $(x(t) * y(t))' = x(t)' * y(t)$.}^,\footnote{ Recall the sifting property of the Dirac delta function $\delta(t)$ implies that $\delta(t) * h(t) = h(t)$.}
\end{equation}
Thus, as described above mathematically, the derivative of the histogram at time $t < 0$ should directly give the detection system timing response function $g(t)$.
In Figure \ref{fig:DetectionResponseFunction}a) the raw $HH + VV$ coincidence histogram is plotted against the time delay $t$ and the red curve is the raw data filtered by a Savitzki-Golay filter to reduce the impact noise.
This filtered coincidence curve of the raw curve is represented by $f(t)$ in the above model. 
The derivative of $f(t)$ is displayed in Figure \ref{fig:DetectionResponseFunction}b).
We call the points where $t \leq 0$ is displayed in red, $\Tilde{g}(t)$, which fully describes the detection system timing response.
Given that the impulse response of each SPAD follows the same function (see Ref. \cite{Fognini2019-uj}), the timing response for the cross-correlation histogram should be a symmetric function.
Thus, the raw data $\Tilde{g}(t)$ is mirrored across the $t = 0$ axis to give $g(t)$ in terms of raw data points.
Then, these points are fitted with the function
\begin{equation} \label{eq:timing-response}
    g_f(t,t_0,A,\sigma) = \frac{A}{\left( \cos{ \left( \frac{t - t_0}{\sigma} \right) } \right)^2} 
\end{equation}
in order to extract the parameters for $g(t)$ to be used in the simulations.
We find the FWHM to be $\sigma = 488 \pm 1$\,
ps.
The $R^2$-value for this fit is 0.9971.
The fitted curve now represents $g(t)$ --- the timing response for our coincidence detection system when using SPADs.
For the SNSPDs, the method discussed above is difficult to use because the system now has very high temporal precision. Thus, there are only a few points around the slope at $t = 0$ making the derivative of the coincidence histogram an unreliable method for finding $g(t)$.  
Instead, we estimate the timing resolution just using the FWHM of the jitter histograms for the SNSPDs and the time tagger stated by the manufacturer, and then assume all timing response functions are Gaussian. See Methods in the main body of the article.

\subsection{Time-resolved Entanglement Fidelity} \label{supp-sec:time-resolved_ent_fidelity}
\begin{figure}[ht]
    \centering
    \includegraphics[width = 0.93\columnwidth]{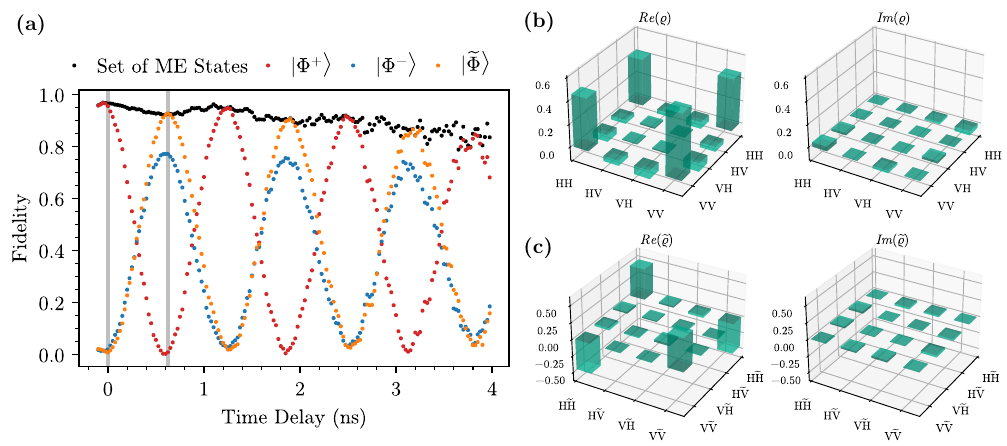}
    \caption{\textbf{Two-photon state as a function of the exciton lifetime.} a) Fidelity of reconstructed density matrix to the set of maximally entangled two-qubit state (black), the $\ket{\Phi^+}$ Bell state (red), the $\ket{\Phi^-}$ Bell state (blue) and the closest maximally entangled pure state at $\tau = 0.625$ (orange) as a function of time delay $\tau$. b) Reconstructed density matrix at $\tau = 0$ ns. c) Reconstructed density matrix at $\tau = 0.625$ ns after rotating qubit one by $U_1(\theta_1,\phi_1)$ and qubit two by $U_2(\theta_2,\phi_2)$ such that it closely resembles the state $\ket{\Phi^-}$.}
    \label{fig:fid_osc_dm_plots}
\end{figure}

Aside from the concurrence, we also calculated the time-resolved fidelity to the set of maximally entangled states ($F_{max}$) \cite{PhysRevA.66.022307} and the fidelity with respect to a particular Bell state, displayed in Figure \ref{fig:fid_osc_dm_plots}a.
The effect of the FSS becomes clear when we plot the fidelity with respect to $\ket{\Phi^+} = \left( \ket{HH} + \ket{VV}\right)/\sqrt{2}$ in Figure \ref{fig:fid_osc_dm_plots}a. 
As per Eq. \ref{eq:bellstate-fss} in the main body, the pair emitted by the QD should start in the $\ket{\Phi^+}$ state and return to the $\ket{\Phi^+}$ state after $\tau = T_S$.
This is precisely the behaviour that we observed, as shown in Figure \ref{fig:fid_osc_dm_plots}a. 
We see very clear, high-visibility oscillations with the fidelity to $\ket{\Phi^+}$ remaining above 80$\%$ up to 4\,ns, at which point more than 99\% of photon pairs have been emitted.

Just as with the concurrence we see small oscillations in $F_{max}$ that result in a reduction of the maximum fidelity from its initial peak at $\tau = 0$ of $96.6 \pm 0.2 \%$ to a minimum at $92.6\pm 0.6\%$ at $\tau = 0.625$\,ns.
However, just like the concurrence, the fidelity comes back to $95.0 \pm 0.8\%$ at $\tau = 1.25$\,ns which is nearly within $1\%$ of the initial peak value. 
Interestingly, we see that the fidelity to the $\ket{\Phi^-} = \left( \ket{HH} - \ket{VV}\right)/\sqrt{2}$ state reaches a maximum of $77\%$ at halfway point of the oscillation period ($\tau = 0.625$\,ns). 
If instead, we plot fidelity to the closest maximally entangled pure state at $\tau = 0.625$\,ns, denoted as $|{\Tilde{\Phi}}\rangle$, we see that it reaches the black dots exactly.
This implies the state is not oscillating exactly as $\left( \ket{HH} + \exp(i\theta)\ket{VV}\right)/\sqrt{2}$ but rather in a rotated version of this state. 
Indeed, if we apply a different unitary transformation to each qubit individually $U_1(\theta_1,\phi_1) \otimes U_2(\theta_2,\phi_2)$, we can transform the state $|\tilde{\Phi}\rangle$ to the state $\ket{\Phi^-}$.
The unitary transformation we apply is in the form of a general waveplate, with parameters $\theta_1 = 0.722, \phi_1 = 1.956$ radians and $\theta_2 = 1.111, \phi_2 = 2.142$ radians \cite{Theocaris1979}.

In Figure \ref{fig:fid_osc_dm_plots}b we plot the reconstructed density matrix at $\tau = 0$.
The density matrix exhibits exactly the characteristics of the $\ket{\Phi^+}$ Bell state, namely negligible imaginary components and four positive real components.
Similarly, in Figure \ref{fig:fid_osc_dm_plots}c we plot the density matrix of the rotated $|\tilde{\Phi}\rangle$ state, namely $\left[ U(\theta,\phi) \otimes U(\theta,\phi) \right] |\tilde{\Phi}\rangle$.
We see what is expected from the $\ket{\Phi^-}$ state: negligible imaginary components; and, two positive real components and two negative real components.
The form of the density matrix in Figure \ref{fig:fid_osc_dm_plots}c confirms that we are indeed able to rotate the initially reconstructed state $|\tilde{\Phi}\rangle$ into the expected $\ket{\Phi^-}$ state.

\subsection{Dephasing-Free Model}\label{supp_sec:dephasing_free_model}
The  model used here is based on the work of Ref. \cite{Fognini2019-uj} which estimates the two-photon density matrix reconstructed from performing quantum state tomography on a QD emitter. 
In this particular model, we assume that the QD emitter is in the ``dephasing-free'' regime, i.e., the two-photon entangled state emitted by the QD is in the maximally entangled state $\ket{\psi(\tau)} = \left(\ket{HH} + \exp(i S \tau / \hbar) \ket{VV} \right)/\sqrt{2}$ where $S$ is the finite exciton fine structure splitting (FSS).
As a result, the probability of detecting a coincidence in a particular two-photon polarization state $\ket{ij} \, i,j \in \{H,V,D,A,R,L\}$ at a given time delay is
\begin{equation}
    p_{ij} (\tau) = |\bra{ij}\ket{\psi(\tau)}|^2 \, \frac{1}{\tau_X} e^{-\tau/\tau_X} \, \Theta(\tau) 
\end{equation}
where $\tau_X$ is the radiative lifetime of the exciton and $\Theta(\tau)$ is the Heaviside step function.
To estimate the total number counts in each histogram bin after an experiment we use 
\begin{equation}
    N_{0} = N_X \, N_{XX} \, \frac{ T_{\textnormal{bin}} \, \Delta \tau }{ f_{\textnormal{rep}} }
\end{equation}
where $N_X$, $N_{XX}$ are the measured count rates of the exciton and biexciton, respectively, $T_{\textnormal{exp}}$ is the experiment runtime, and $f_{\textnormal{rep}}$ is the excitation laser repetition rate.
However, even though the model assumes the QD emitter experiences no dephasing, we still include the non-zero multiphoton emission probability of the QD emitter, and the finite dark count and timing resolution of the single-photon detectors used in the experiment.

To incorporate the effect of detector dark counts which are unpolarized and therefore add unwanted coincidences to all polarization basis measurements.
We estimate the expected number of coincidences from dark counts with the equation
\begin{equation}
    N_{d}^{cc} \simeq \Big[ N_{XX} \, N_{d_X} + N_{X} \, N_{d_{XX}} \Big] \, \frac{ T_{\textnormal{bin}} \, \Delta \tau }{ f_{\textnormal{rep}} }
\end{equation}
where $N_{d_X}$ and $N_{d_{XX}}$ are the dark count rates for the single photon detectors measuring exciton and biexciton photons, respectively.
The finite timing resolution of the detection system is included by convolving the time-dependent count probability distribution with the detection system response function $g(\tau)$.
The coincidence count histogram versus time delay for a given polarization basis measurement $ij$ is thus
\begin{equation}
    N_{ij} (\tau) = \left( N_0\,p_{ij}(\tau) + N_d \right) * g(\tau) \, \Delta \tau
\end{equation}
where $*$ is the convolution operation. 
With the simulated $N_{ij}(\tau)$ coincidence histogram, the density matrix can be reconstructed in 50\,ps width time bins to give $\rho(\tau)_{QD}$.
Finally, the effect of multi-photon emission on the reconstructed density matrix is included by adding the 4x4 identity matrix $I/4$, i.e., uncorrelated light.
The final simulated density matrix is then 
\begin{equation}
    \rho_{\textnormal{sim}}(\tau) = (1 - p_m) \, \rho(\tau)_{QD} + p_m \, \frac{I}{4}
\end{equation}
where $p_m \simeq ( \, \tilde{g}^{(2)}_{XX}(0) + \tilde{g}^{(2)}_{X}(0) \, )/2$ \cite{Neuwirth2022-sx}.
Time slices of $\rho_{\textnormal{sim}}(\tau)$ with a width of 50\,ps, were then used to calculate the time-resolved concurrence curves for the SPAD and SNSPDs displayed in Figure \ref{fig:figure3}.

\subsection{Polarized HBT Measurement}
\begin{figure}[H]
    \centering
    \includegraphics[width = 0.9\textwidth]{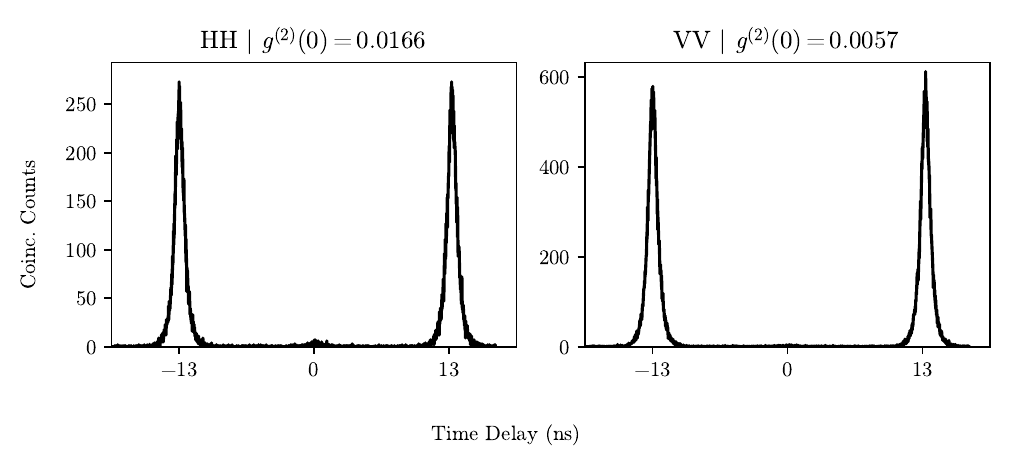}
    \caption{Hanburry Brown Twiss autocorrelation histograms taken just after performing quantum state tomography, when the emission from the quantum dot is polarized to either H or V. We see a significantly higher number of coincidences at $t = 0$ in the H polarized histogram, which is likely due to stray photons that were not successfully filtered. }
    \label{fig:polarized_hbt}
\end{figure}

\subsection{Dephasing Model}\label{supp_sec:dephasing_model}
\begin{figure}[ht!]
    \centering
    \includegraphics[width = \textwidth]{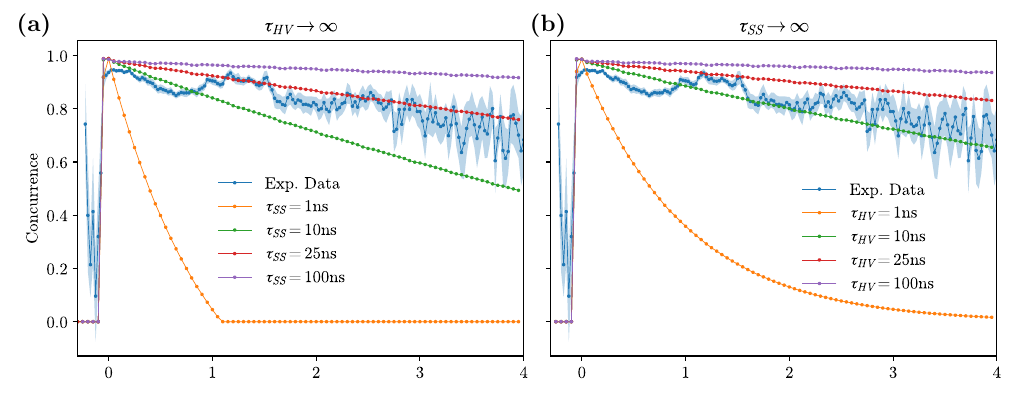}
    \caption{Time-resolved concurrence curve from the dephasing model for different characteristic interaction times. a) In the limit of no cross-dephasing, the time-resolved concurrence is calculated for different spin-scattering interaction times. b) In the limit of no spin-scattering, the time-resolved concurrence is calculated for different cross-dephasing interaction times. }
    \label{fig:dephasing}
\end{figure}

To explore the effects that a dephasing mechanism would have on the experimental time-resolved concurrence curve, we combined the model described in Ref. \cite{Rota2020-wc} with the dephasing free-model outline in the Supplementary Note \ref{supp_sec:dephasing_free_model}.
The model incorporates two types of spin-dephasing: 1) spin-flips (i.e., spin-scattering) of the exciton spin; and 2) cross-dephasing between the H-polarized and V-polarized exciton spin \cite{Hudson2007-di}.
Mathematically, spin-flips are modelled as the addition of uncorrelated light (i.e., the identity) over an interaction period of $\tau_{SS}$, and cross-dephasing as the addition of $\rho_{HV} = \left( \ket{HH}\bra{HH} + \ket{VV}\bra{VV} \right)/2$ (i.e., the loss of coherence between $\ket{HH}$ and $\ket{VV}$) with an interaction period of $\tau_{HV}$.
The final simulated density matrix when dephasing is included is written as,
\begin{equation}
\begin{split}
    \rho_{\textnormal{sim}}(\tau) = k \, e^{-\tau/\tau_{SS}}e^{-\tau/\tau_{HV}}\rho(\tau)_{QD} + \left( 1 - k \, e^{-\tau/\tau_{SS}} \right) \frac{I}{4} \, + k 
    \, e^{-\tau/\tau_{SS}}(1 - e^{-\tau/\tau_{HV}}) \, \rho_{HV}
\end{split}
\end{equation}
where $k = 1 - p_m$.
With this model for spin-dephasing, we can now quantify the effect of the detection system (included in $\rho(\tau)_{QD}$) and multi-photon emission (incorporated through $k$) plus the effect of spin-scattering and cross-dephasing. 
We explore the effect of both spin-flip and cross-dephasing on the time-resolved concurrence by looking at each mechanism individually and comparing it to the concurrence curve from the experimental data (see Figure \ref{fig:dephasing} in Sec \ref{supp_sec:dephasing_model}).
Specifically, two distinct limits are examined: varying the spin-flip coherence time $\tau_{SS}$ with no cross-dephasing ($\tau_{HV} \rightarrow \infty$); and varying the cross-dephasing time $\tau_{HV}$ with no spin-flip dephasing ($\tau_{SS} \rightarrow \infty$).
As expected, the curves displayed in Figure \ref{fig:dephasing} do not account for the dip in the concurrence between $\tau = 0$\,ns and $\tau = 1.3$\,ns, implying that it is not spin-scattering and cross-dephasing that are responsible.
However, the model indicates that dephasing times about an order of magnitude larger than the lifetime could explain the slow reductions in the experimental concurrence when $\tau > 2 $\,ns.
Still, these dephasing values are much greater than the radiative lifetime, implying that their impact will not degrade the lifetime-average concurrence, and therefore the source could still be considered ``dephasing-free''.

\newcommand{\deltaleak}{\delta_{\text{leak}}}
\renewcommand{\Tr}[1]{\textrm{Tr}\left [ #1 \right ]}        %
\newcommand{\ntimebin}{{n_{\textrm{bin}}}}
\newcommand{\nrounds}{{n_{\textrm{round}}}}
\newcommand{\vecA}{{\vec{i}}} %
\newcommand{\vecB}{{\vec{j}}} %
\newcommand{\vecAn}{{\vec{n}}} %
\newcommand{\vecBn}{{\vec{m}}} %
\newcommand{\vecASingle}[1]{{\vec{i}_{#1}}} %
\newcommand{\vecBSingle}[1]{{\vec{j}_{#1}}} %

\subsection{Using quantum dots with FSS for QKD} \label{supp:QKD}

In this supplement, we describe a general method to calculate secret key rates for entanglement-based (EB) QKD experiments that use fast detectors. The idea is to break each round of the protocol into $\ntimebin$ time-bins where the detectors record the click patterns in each time-bin. We then use this to calculate key rates for the QD source with non-zero FSS.

In our security proof, we assume that the threshold detectors used have no memory effects (detector dead times, afterpulsing, etc.), and all detectors have the same detection efficiency.
Any security proof for independent and identically distributed (iid) states can be lifted to a security proof for general states via the quantum de Finetti theorem \cite{renner2007symmetry} or the postselection technique \cite{christandl2009postselection}, which are asymptotically tight. 
As we only consider asymptotic key rates, we can thus without loss of generality assume that Alice and Bob's shared state is iid i.e. $\rho_{A^{\nrounds}B^{\nrounds}} = \rho_{AB}^{\otimes \nrounds}$, where $\nrounds$ is a large number of protocol rounds. In the rest of this supplement we focus on proving security for these iid states.

\subsubsection{QKD Protocol Steps}

Here we outline the steps in a generic time-resolved entanglement-based (EB) protocol with $\nrounds$ rounds.
\begin{enumerate}
    \item \label{QKDprotocol:transm} \textbf{Signal Transmission:} Eve prepares optical states $\rho_{AB}$ to send to Alice and Bob.
    \item \textbf{Measurement:} Alice and Bob each have sets of POVM elements $\{\Gamma^{A}_{\vecA}\}$ and $\{\Gamma^{B}_{\vecB}\}$. Here, the elements $i_{k}$ and $j_{k}$ of the vectors $\vecA$ and $\vecB$ represent the click patterns observed in time-bin $k$.
\end{enumerate}
After repeating the above steps $\nrounds$ times, we proceed to the next part of the protocol.
\begin{enumerate}[resume]
    \item \label{QKDprotocol:accTest} \textbf{Acceptance Testing:} Alice and Bob randomly choose a subset of the rounds for testing. For the rounds chosen for testing, they both publicly announce their measurement results $\vecA$ and $\vecB$ to form a distribution of the frequency of events. They then check if this frequency distribution belongs to the acceptance set agreed upon before running the protocol. If it does, they proceed with the protocol after discarding the test results. Otherwise, they abort.
    \item \label{QKDprotocol:announce} \textbf{Announcements and Sifting:} Alice and Bob make announcements over the authenticated classical channel. They sift the non-tested data based on the announcements made i.e. they choose a subset of signal and measurement data to keep and discard the rest based on the announcements.
    \item \label{QKDprotocol:keyMap} \textbf{Key Map:} Alice uses her measurement results $\vecA$ as well as the announcements to map her data into a key string $x$. This is called the raw key.
    We assume here that the key is a bit string for simplicity, but all the steps can be applied more generally.
    \item \label{QKDprotocol:protocolStepErrorCorr} \textbf{Error Correction:} Alice and Bob then perform error correction over the authenticated classical channel to make Bob's measurement outcomes $\vecB$ match with Alice's bit string $x$. We denote the data communicated per key bit to Eve in this process as $\deltaleak$.
    \item \textbf{Privacy Amplification:} Alice and Bob produce their final secret key by applying an appropriate hash function on the raw key (Theorem 5.5.1 of \cite{renner2008security}).
\end{enumerate}

\subsubsection{Key rate calculation}

\begin{itemize}
    \item The secret key rate in the asymptotic limit in the iid case can be found using the Devetak-Winter formula \cite{devetak2005distillation}: $R = H(Z\vert E)-\deltaleak$ where $Z$ is the key register, $E$ is Eve's register, and $\deltaleak$ is the number of bits per round leaked to Eve during Step \ref{QKDprotocol:protocolStepErrorCorr} of the protocol.
    \begin{itemize}
            \item A lower bound for the key rate can be found by minimising $H(Z\vert E)$ over Eve's marginal states consistent with the observed statistics.
            \item Following \cite{coles2016numerical,winick2018reliable}, this minimization can be reformulated as an SDP
            \begin{equation} \label{eq:mainKeyRate}
                \begin{aligned}
                    R = \underset{\rho_{AB}}{\textrm{min }} &D\left(\mathcal{G}\left(\rho_{AB}\right)\vert\vert\mathcal{Z}\left(\mathcal{G}\left(\rho_{AB}\right)\right)\right) - \delta_\textrm{leak} \\
                    \textrm{s.t. } &\Tr{\left(\Gamma_{\vecA}\otimes \Gamma_{\vecB}\right)\rho_{AB}} = p_{\vecA \vecB} \quad\quad\forall \vecA,\vecB\\
                    &\rho_{AB} \geq 0,
                \end{aligned}
            \end{equation}
            where $A$ and $B$ are Alice and Bob's registers. The observed statistics $p_{\vecA\vecB}$ is the probability of Alice and Bob observing outcomes $\vecA$ and $\vecB$ respectively. Note that in phrasing the optimization problem in this way, we have implicitly chosen a point acceptance set (Step \ref{QKDprotocol:accTest} of protocol steps) to be exactly the statistics from the honest implementation. This choice is meaningful as we work in the asymptotic limit.
            \item Here, the relative entropy $D\left(\mathcal{G}\left(\rho_{AB}\right)\vert\vert\mathcal{Z}\left(\mathcal{G}\left(\rho_{AB}\right)\right)\right)-\deltaleak$ is the objective function where $\mathcal{G}$ is a map that represents the protocol (including announcements), and $\mathcal{Z}$ is a map that can be constructed from the key map. Since we do not use most of the specific details of these maps, we abstract the objective function as $f(\rho_{AB})$. For more details, see \cite{winick2018reliable, coles2016numerical}.
        \end{itemize}
        \item The aim is to reduce this to a problem that is numerically tractable. There are two obstacles to doing this.
        \begin{itemize}
            \item The first is that each time-bin has an infinite-dimensional Fock space associated with it. This is a well-studied problem in QKD \cite{fung2011universal,gittsovich2014squashing,li2020improving,upadhyaya2021dimension}. Any of these methods can be used to reduce the problem to solving finite-dimensional SDPs in each time-bin. 
            \item The dimension of the space after this process is still large due to the multiple time-bins. Thus, we will simplify Eq. (\ref{eq:mainKeyRate}) to a minimization on each time-bin individually to obtain a ``time-resolved key rate''.
        \end{itemize}
    \end{itemize}

    \subsubsection{Time-resolved key rate}
    
    \begin{itemize}
        \item Since we assume that the threshold detectors have no memory effects, each of Alice and Bob's POVM elements can be modeled as $\Gamma_{\vec{i}}=\bigotimes_{k=1}^\ntimebin\Gamma_{i_k}$.
        \begin{itemize}
            \item For threshold detectors such as the SNSPDs, $\Gamma_{i_k}$ is block diagonal in the total number of photons in time-bin $k$ i.e.
            \begin{equation}
                \Gamma_{i_k}=\bigoplus_{n_k=0}^\infty\Gamma_{i_k}^{n_k}.
            \end{equation}
            \item Thus, the POVM elements $\Gamma_{\vec{i}}$ are block-diagonal as
            \begin{equation}
                \Gamma_{\vec{i}} = \bigoplus_{\vec{n}}  \Gamma^{n_1}_{i_1}\otimes  \Gamma^{n_2}_{i_2} \otimes \ldots \otimes \Gamma^{n_\ntimebin}_{i_\ntimebin}.
            \end{equation}
            \item As a result, we can without loss of generality (see for instance, Observation 8 from \cite{gittsovich2014squashing}) assume that Alice and Bob's joint state has the same block-diagonal structure
            \begin{equation}
                \rho_{AB} = \bigoplus_{\vecAn, \vecBn} \rho_{AB}^{\vecAn\vecBn},
            \end{equation}
            where $\rho_{AB}^{\vecAn\vecBn}$ are sub-normalized states.
        \end{itemize}
        \item Following Eq. (D.9) from \cite{li2020application}, we can exploit the block-diagonal structure in the state to write $f(\rho_{AB}) = \sum_{\vecAn,\vecBn} f(\rho_{AB}^{\vecAn\vecBn})$, where $f(\rho_{AB})$ is the objective function of the SDP in Eq. (\ref{eq:mainKeyRate}). Thus, $R\geq \sum_{\vecAn,\vecBn} R^{\vecAn\vecBn}$ where 
        \begin{equation} \label{eq:BlockKeyRate}
            \begin{aligned}
                R^{\vecAn\vecBn} = \underset{\rho_{AB}}{\textrm{min }} &f\left(\rho_{AB}^{\vecAn\vecBn}\right)\\
                \textrm{s.t. } &\Tr{\left(\Gamma_{\vecA}\otimes \Gamma_{\vecB}\right)\rho_{AB}} = p_{\vecA \vecB} \quad\quad\forall \vecA,\vecB\\
                &\rho_{AB} \geq 0\\
                &\rho_{AB} = \bigoplus_{\vecAn, \vecBn} \rho_{AB}^{\vecAn\vecBn}.
            \end{aligned}
        \end{equation}
        Since each $R^{\vecAn\vecBn}$ is positive, taking finitely many of these terms is sufficient to lower bound the key rate. In particular, we consider those terms where all but one time-bin has no photons on each of Alice and Bob's sides.
        \begin{itemize}
            \item Note that $\rho_{AB}^{\vecAn_{k_c}\vecBn_{l_c}}=\rho_{AB}^{n_{k_c}m_{l_c}}\bigotimes_{k\neq k_c, l\neq l_c}\ketbra{0}_k\otimes\ketbra{0}_l$ where $k_c$ and $l_c$ are the time-bins with photons on Alice and Bob's side respectively.
        \end{itemize}
        \item If the detection efficiency of all detectors is identical, it is sufficient to prove security for lossless detectors instead \cite{lutkenhaus1999estimates} by considering loss to be a part of the channel. In this case, for all Alice's events $\vecASingle{k_c}$ that have detection events in just time-bin $k_c$, 
        $\Gamma_{\vecASingle{k_c}} = \Gamma_{i_{k_{c}}}\bigotimes_{k\neq k_{c}} \ketbra{0}_{k}$. Similarly, $\Gamma_{\vecBSingle{l_c}} = \Gamma_{j_{l_{c}}}\bigotimes_{l\neq l_{c}} \ketbra{0}_{l}$ for all Bob's events $\vecBSingle{l_c}$ that have detection events in just time-bin $l_c$.
        Thus, 
        \begin{align}
            \nonumber \Tr{\left(\Gamma_{\vecASingle{k_c}}\otimes \Gamma_{\vecBSingle{l_c}}\right)\rho_{AB}} &= \Tr{\left(\Gamma_{\vecASingle{k_c}}\otimes \Gamma_{\vecBSingle{l_c}}\right)\sum_{n_{k_c},m_{l_c}=0}^\infty\rho_{AB}^{\vecAn_{k_c}\vecBn_{l_c}}}\\
            &= \Tr{\left(\Gamma_{i_{k_c}}\otimes \Gamma_{j_{l_c}}\right)\rho_{AB}^{{k_c}{l_c}}},\label{eq:QKDDetectorSimpllify}
        \end{align}
        where $\vecAn_{k_c}$ ($\vecBn_{l_c}$) are vectors where all but the $k_c$ ($l_c$) entry is 0, and $\rho_{AB}^{{k_c}{l_c}} = \sum_{n_{k_c},m_{l_c}=0}^\infty \rho_{AB}^{n_{k_c}m_{l_c}}$. Note that the right hand side of Eq. (\ref{eq:QKDDetectorSimpllify}) is described in terms of operators that act on the space for a single time-bin $k_c$, $l_c$ for Alice and Bob respectively.
        \begin{itemize}
            \item For all events $\vecA$, $\vecB$ with clicks in more than one time-bin, $\Tr{\left(\Gamma_{\vecA}\otimes \Gamma_{\vecB}\right)\rho_{AB}^{\vecAn_{k_c}\vecBn_{l_c}}} = 0$.
        \end{itemize}
        \item Additionally, we define the function $g\left(\rho_{AB}^{{k_c}{l_c}}\right) \coloneqq \sum_{n_{k_c},m_{l_c}=0}^\infty f\left(\rho_{AB}^{\vecAn_{k_c}\vecBn_{l_c}}\right)$ acting on the space for a single time bin for Alice and Bob.
        Thus, the SDP in Eq. (\ref{eq:BlockKeyRate}) can be simplified as
        \begin{equation} \label{eq:timeResolvedKeyRate}
            \begin{aligned}
                R^{k_c l_c} = \underset{\rho_{AB}^{{k_c}{l_c}}}{\textrm{min }} &g\left(\rho_{AB}^{{k_c}{l_c}}\right)\\
                \textrm{s.t. } &\Tr{\left(\Gamma_{i_{k_c}}\otimes \Gamma_{j_{l_c}}\right)\rho_{AB}^{{k_c}{l_c}}} = p_{\vecASingle{k_c} \vecBSingle{l_c}} \quad\quad\forall \vecASingle{k_c},\vecBSingle{l_c}\\
                &\rho_{AB}^{{k_c}{l_c}} \geq 0.
            \end{aligned}
        \end{equation}
        \item We define $r(\tau)=\frac{\sum_{\abs{k_c-l_c}=\tau}R^{k_c l_c}}{\Tr{\rho_{AB}^{\tau}}}$ to be the time-resolved key rate where $\rho_{AB}^{\tau} = \sum_{\abs{k_c-l_c}=\tau} \rho_{AB}^{{k_c}{l_c}}$.
        Thus, we have simplified the key rate SDP in Eq. (\ref{eq:mainKeyRate}) to the time-resolved key rate SDPs described in Eq. (\ref{eq:timeResolvedKeyRate}) which act on the spaces corresponding to a single time-bin $k_c$, $l_c$ for Alice and Bob respectively.
        \item The time-resolved key rate SDPs described in Eq. (\ref{eq:timeResolvedKeyRate}) are still infinite-dimensional. As discussed above, any of the techniques from \cite{fung2011universal,gittsovich2014squashing,li2020improving,upadhyaya2021dimension} to reduce this to a finite-dimensional optimization can be used to obtain lower bounds on the time-resolved key rate. For the purposes of the example we consider in Section \ref{appSec:QKDPoPExp}, we use the dimension reduction method from \cite{upadhyaya2021dimension}.
\end{itemize}

\subsubsection{Proof-of-principle experiment} \label{appSec:QKDPoPExp}

So far, we have described a generic framework to compute key rates with fast detectors i.e. detectors with low timing jitter and dead times to resolve subsequent time-bins with independent detection events. We now describe how we used this framework to obtain the plots in Figure \ref{fig:qkd_keyrate}, which clearly demonstrate the utility of a QD with non-zero FSS. Note that in the proof-of-principle experiment, we consider the case where only single-click events, i.e. a click only in a single detector time-bin in each protocol round for Alice and Bob each. As we expect our source to primarily send out single-photon states to Alice and Bob each, this would be close to what we expect to observe in a real experiment.
\begin{itemize}
    \item We consider a polarization-based 6-state protocol \cite{bruss1998optimal} where both Alice and Bob measure their state in 3 conjugate polarization bases which we term the `X', `Y' and `Z' bases. A schematic of the setup is shown in Figure \ref{fig:time_resolved_qkd_schematic}. We set the probability of measuring in the Z-basis to be $\sqrt{0.99}$ each. After testing, they announce their basis data and discard all signals not in the Z-basis. Alice then maps single-clicks in the Z-basis into bits. Finally, they do error correction and privacy amplification.
    \begin{figure}
        \centering
        \includegraphics{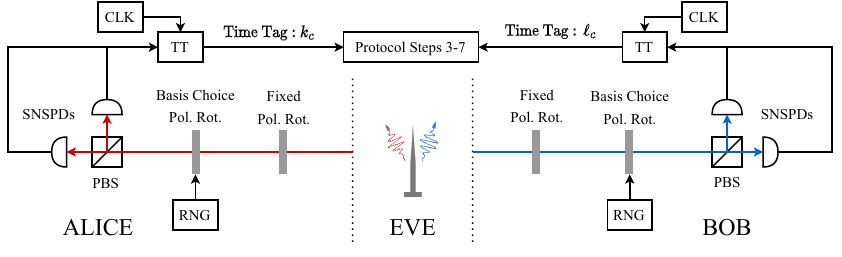}
        \caption{Schematics of proof-of-principle experimental setup. PBS: polarisation beamsplitter; RNG: random number generator; SNSPD: superconducting nanowire single photon detector; TT: time tagger; CLK: clock.
        The basis choice polarisation rotator together with the RNG randomly choose between three mutually unbiased bases to measure in. The choice of which three mutually unbiased bases are used for the protocol is made by the fixed polarisation rotator. The clock rate must be faster than the FSS oscillations of the quantum dot.}
        \label{fig:time_resolved_qkd_schematic}
    \end{figure}
    
    \item For the key rate calculation, we simulate the observations we expect to see from the characterized nanowire QD entangled photon source.
    \begin{itemize}
        \item We use the dimensional reduction method \cite{upadhyaya2021dimension} to construct finite-dimensional SDPs from Eq. (\ref{eq:timeResolvedKeyRate}) which allows us to consider projections of measurments and states in a subspace only. For this, we need the weight $W = \Tr{\Pi_1 \rho^{k_c l_c}\Pi_1}$ outside the projected subspace. Thus, the time-resolved key rate can be lower bounded by the solution to the following finite-dimensional SDP:
        \begin{equation}
            \begin{aligned}
                R^{k_c l_c} \geq \underset{\rho_{AB}^{11}}{\textrm{min }} &g\left(\rho_{AB}^{11}\right)\\
                \textrm{s.t. } &p_{\vecASingle{k_c} \vecBSingle{l_c}}-W\leq \Tr{\left(\Gamma_{i_{k_c}}^1\otimes \Gamma_{j_{l_c}}^1\right)\rho_{AB}^{11}} \leq p_{\vecASingle{k_c} \vecBSingle{l_c}} \quad\quad\forall \vecASingle{k_c},\vecBSingle{l_c}\\
                &\rho_{AB}^{11} \geq 0,
            \end{aligned}
        \end{equation}
        where $\Gamma^1 = \Pi_1 \Gamma \Pi_1$ is the projection of the POVM elements into the single-photon subspace $\Pi_1$. Note that since the POVM elements commute with our chosen projection, we have used the tighter Eq. (49) from \cite{upadhyaya2021dimension}.
    \end{itemize}
    Here, the weight outside the subspace $W$ can be estimated from the multi-click events (see for e.g.- \cite{narasimhachar2011study,li2020improving, nahar2023imperfect} for some protocol-dependent techniques to do this). In our case, since we have assumed to observe only single-click events, this can be calculated exactly to be $W=0$. Thus, the time-resolved key rate SDP simplifies to numerically optimising an SDP in the qubit space.
    \item Additionally since we are performing the 6-state experiment, we have a tomographically complete set of measurements on the qubit space. Thus, the state $\rho_{AB}^{11}$ can be determined exactly from the observations for each time-bin. As a result, we can compute the time-resolved key rates $r(\tau)$ by simply computing the objective functions for the qubit density matrix for the time-bins. So no SDP optimizations are required for our simple example.
    \item Finally, as described in Supplementary Note \ref{supp-sec:time-resolved_ent_fidelity}, the actual entangled photon pair does not rotate in the $HH-VV$ plane as described by Eq. (\ref{eq:bellstate-fss}). Thus, our initial key map consisting of $H$ and $V$ polarizations might not be optimal. We instead run an optimization over all choices of conjugate polarization bases to measure in. 
    Specifically, just as in Supplementary Note \ref{supp-sec:time-resolved_ent_fidelity}, we apply two independent unitary transformations $U_1(\theta_1,\phi_1) \otimes U_2(\theta_2,\phi_2)$ to the reconstructed two-photon state, and find the set of parameters  that maximizes the key rate $R$.
    We use the result of this optimization to create the blue curve in Figure \ref{fig:qkd_keyrate}. Practically, this can be implemented in the protocol by adding appropriate waveplates before Alice and Bob's entire measurement setup as depicted in Figure \ref{fig:time_resolved_qkd_schematic}.
\end{itemize}

\subsection{Blinking}\label{supp_sec:blinking}
For a QD emitter to be considered truly ``on-demand'', a photon should be emitted for every incident excitation pulse. 
Consequently, if excitation pulses are incident at a regular, periodic interval $T_{\textnormal{rep}}$ on the QD, then photons should be emitted in a matching temporal pulse train pattern.
Thus, the coincidence histogram from the HBT autocorrelation experiment will consist of coincidence peaks every $T_{\textnormal{rep}}$ except near the $\tau \rightarrow 0$ as a true single photon source should exhibit strong anti-bunching. 
The height of the histogram side peaks at $\tau \neq 0$ represents the probability of detecting a photon at some time delay $\tau$ after an initial photon has been detected \cite{Santori2004-rv}.
Thus for a true on-demand source, the heights of these $\tau \neq 0$ peaks should all be equal, as there is an equal probability of a coincidence occurring between any photons emitted in the train.
This situation where all peaks are of equal height is known as the ``Poisson'' limit, with the corresponding peak heights known as the ``Poisson'' level \cite{Zhai2020-yh}.
Typically, the phenomenon of luminescence blinking leads to a deviation from these flat peaks for all $\tau \neq 0$.
Instead, the height of the peaks in the histogram will increase as the time delay $\tau$ decreases.
This means that, given the detection of an initial photon emitted from the QD, there is a higher probability of detecting a photon from the excitation pulse that directly followed, compared to the Poisson limit where photons are equally likely to arrive at any $\tau$ \cite{Santori2004-rv}.
Such behaviour can be shown to result from the QD transitioning between a state that allows for the generation of the biexciton via two-photon resonant excitation, to a state that forbids it and thus forbids the generation of an entangled photon pair. 
In other words, the QD will transition from a biexciton ``ON'' state to a biexciton ``OFF'' state.
For example, if a charge carrier tunnels into an empty QD (ground state $g$) between excitation pulses, then the QD will then be in some charge state $c$.
This extra charge will block the generation of a biexciton via a resonant two-photon absorption, and hence there will be no subsequent emission of an exciton or a biexciton photon \cite{Yang2022-mx}.
In order for optical excitation to occur, the charge must then tunnel out of the QD or recombine with an opposing carrier.
This inherently takes some finite amount of time during which the QD will not be producing any biexciton or exciton photons. 
Hence, a simple random telegraph (ON-OFF) model can be applied to quantify the autocorrelation function that is observed in the presence of luminescence blinking \cite{Jahn2015-eq},
\begin{equation} \label{eq:blinking}
    g^{(2)}(\tau) = 1 + \frac{1 - \beta}{\beta}e^{-\tau/\tau_b} .
\end{equation}
Here, $\beta$ is the probability of the QD being in the ON state at any given time and $1/\tau_b = 1/\tau_{g} + 1/\tau_{c}$ is the transition rate between the two states discussed above, i.e., the rate at which the QD transitions from the ground state to the charge state and then vice versa, i.e.,  ON$\rightarrow$OFF and OFF$\rightarrow$ON, respectively \cite{Yang2022-mx}. 

\begin{figure}[h!]
    \centering
    \includegraphics[width = 0.85\textwidth]{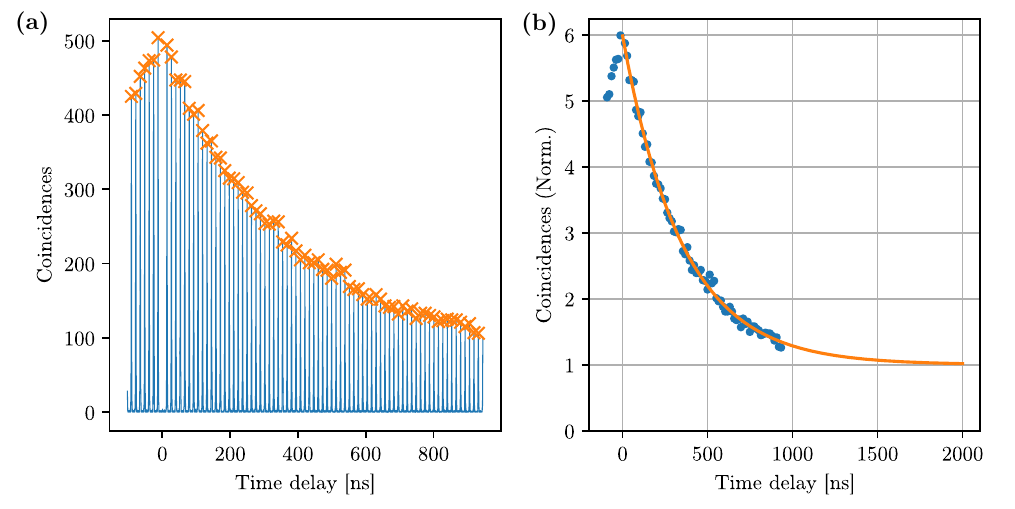}
    \caption{a) Raw data from the Hanburry-Brown-Twiss autocorrelation histogram for the biexciton. The orange x's represent the location of the maximum of each peak. b) Fit of Eq. \ref{eq:blinking} to the peak heights extracted from the plot to the left as a function of time delay $\tau$.}
    \label{fig:blinking}
\end{figure}

In Figure \ref{fig:blinking}, we can see the results of the fit of Eq. \ref{eq:blinking} to the HBT autocorrelation histogram for biexciton emission, from which we extract $\beta_{XX} = 0.167 \pm 0.006$ and $1/\tau_{b,XX} =  2.86 +/- 0.07$\,MHz.
When fitting, we scale the entire Eq. \ref{eq:blinking} function by a factor $A_P$, which represents the peak heights of the raw histogram at the Poisson level, i.e., the peak heights at long $\tau$ where $g^{(2)}(\tau) = 1$.
From the fit, we find the $A_P = 84 \pm 3$ for the biexciton. 
When performing the same fit procedure for the exciton HBT histogram we find $\beta_{X} = 0.186 \pm 0.007$ and $1/\tau_{b,X} = 2.88 +/- 0.09$\,MHz.
Using the extracted $\beta$ values, we can calculate the true  $\tilde{g}^{(2)}(0)$ values, which must be normalized to the Poisson level when blinking is present, with the expression $g^{(2)}(0) = \tilde{g}^{(2)}(0)/\beta$ where $\tilde{g}^{(2)}(0)$ is the value normalized to the nearest neighbour peaks we quoted in the main body.
Doing this, we find $g^{(2)}_X(0) = 0.029 \pm 0.002$ and $g^{(2)}_{XX} = 0.032 \pm 0.002$.
We note the reason for stating the smaller $\tilde{g}^{(2)}(0)$ values in the main body is because it is used to calculate the multi-photon emission probability, which is the relevant metric for quantum state tomography \cite{Neuwirth2022-sx}.

\subsection{Estimating Pair Extraction Efficiency}\label{supp_sec:eff_est}
An estimate for the pair extraction efficiency of the source can be found by examining $\eta_{\textnormal{int}}$ and $\eta_{\textnormal{ext}}$ the ``internal'' and ``external'' efficiencies of the system, respectively. 
Internal efficiency is the probability that a given excitation pulse will generate an exciton and a biexciton.
It is determined by the exciton and biexciton preparation probabilities from the fits to the Rabi oscillations and the blinking ratio: $\eta_{\textnormal{int}} = \eta_{\textnormal{prep},X} \, \eta_{\textnormal{prep},XX} \, \eta_{\textnormal{blink}}$.
The preparation fidelity and blinking ratios are $\eta_{\textnormal{prep},XX} = 0.82 \pm 0.01$, $\eta_{\textnormal{prep},X} = 0.774 \pm 0.009$ and $\eta_{\textnormal{blink}} = 0.167 \pm 0.006$, respectively, and their calculation were presented earlier in Sec. \ref{supp_sec:rabi} and in Sec. \ref{supp_sec:blinking}, respectively.  
Next, we need to estimate the external efficiency of the nanowire, which is the probability that a biexciton-exciton photon pair emitted by the QD will be subsequently coupled out of the nanowire into the propagating free-space optical mode collected by the objective lens.
We approximate this efficiency by using the detected count rates for the exciton and biexciton at saturation under pulsed quasi-resonant excitation (from Ref. \cite{Fognini2019-uj}),
\begin{equation}
    \eta_{\textnormal{NW}} = \frac{N_X N_{XX}}{(\eta_{\textnormal{opt}} \Gamma)^2} = \frac{(942.89 \textnormal{kHz} )(401.29 \textnormal{kHz})}{(0.063)^2 (76.2 \textnormal{MHz})^2} = 0.016  \, .
\end{equation}
The overall predicted efficiency is then $\eta_{est} = \eta_{\textnormal{int}} \, \eta_{\textnormal{ext}} = 0.17 \pm 0.01 \%$, which is within the same order of magnitude as the experimental pair extraction efficiency of $\eta_{\textnormal{exp}} = 0.65 \pm 0.02 \%$.
We note that the estimate of 0.17 is likely an underestimate because we are not including any of the other minor exciton complexes, which appear during quasi-resonant exciton, in the calculation of the nanowire extraction efficiency.
If this modest contribution was included in the extraction efficiency of the nanowire then the two numbers will likely be in closer agreement.  

\subsection{Experimental Apparatus}\label{supp_sec:exp_apparatus}
\begin{figure}[h!]
    \centering
    \includegraphics[width = \textwidth]{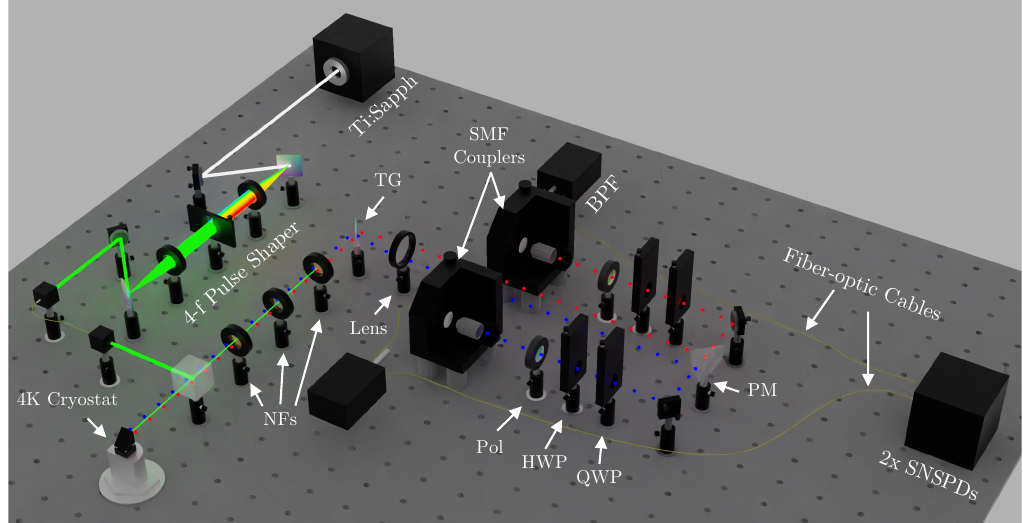}
    \caption{Acronyms: TG --- transmission grating, BPF --- fiber coupled bandpass filter, PM --- right-angle prism mirror, NFs --- notch filters, SNSPDs --- superconducting nanowire single-photon detectors, Pol --- linear polarizer, HWP --- half-waveplate, QWP --- quarter-wave plate. }
    \label{fig:exp_apparatus}
\end{figure}

The experimental setup consisted of three main parts: generating the excitation pulse, quantum state tomography of the photon pairs from the nanowire QD sample, and single-photon detection.
An illustration of the entire experimental apparatus can be seen in Figure \ref{fig:exp_apparatus}. 
See the Methods section for a written description.

\subsection{Apparatus Calibration}\label{supp_sec:apparatus_calibration}
\begin{figure}[h!]
    \centering
    \includegraphics[width = 0.75\textwidth]{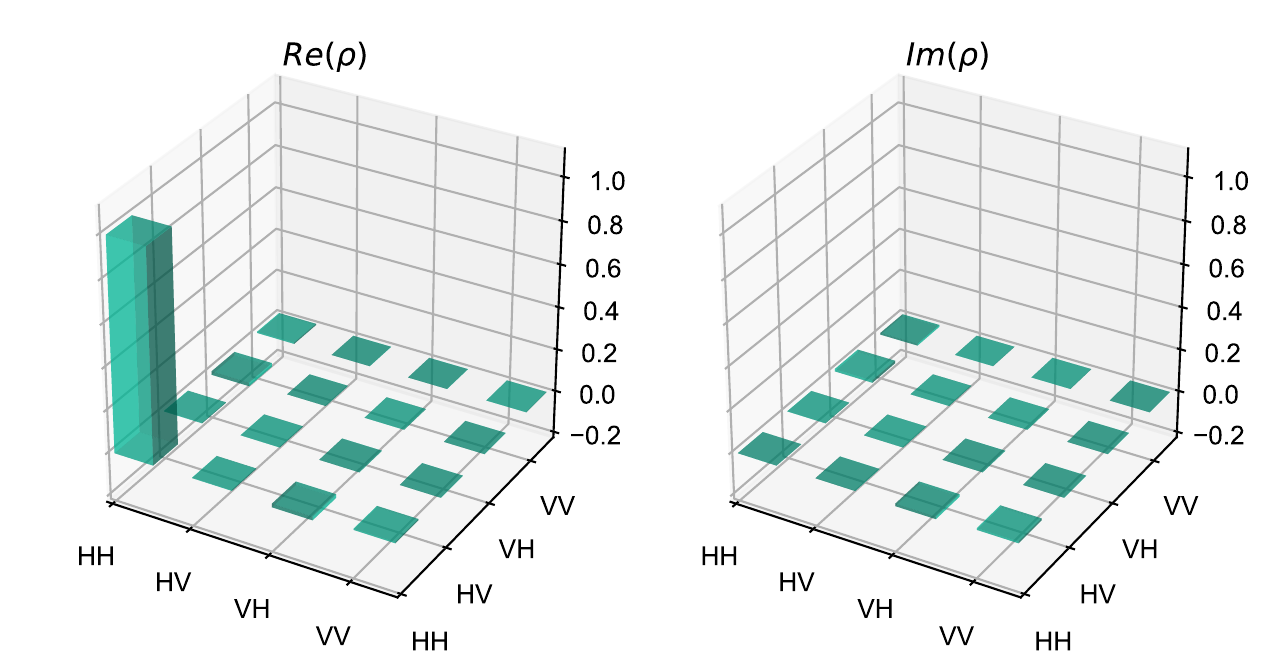}
    \caption{Reconstructed density matrix from performing quantum state tomography on an H-polarized laser signal. The single column in the real part of $\rho$ at $\ket{HH}\bra{HH}$ demonstrates that the system was well calibrated with respect to H polarized photons.}
    \label{fig:calibration_dm}
\end{figure}

To calibrate the experimental setup pictured above, we first found the principles axes of the $\lambda/2$ and $\lambda/4$ waveplates with respect to $H$-polarized light in the lab frame.
This was done by sending in a weak $H$ polarized laser signal through the QD excitation path, from which it reflected off the device substrate and then followed the path of the QD emission through both waveplates and the linear polarizer, which has its polarizing axis fixed at 0 degrees to the horizontal, as illustrated in Figure \ref{fig:exp_apparatus}.
With the $H$-polarized laser running, the photo-intensity coupled into the single-mode fiber was monitored while one waveplate was at a fixed angle and the other was rotated. 
For example, the $\lambda/2$ waveplate rotated, while the $\lambda/4$ waveplate remained fixed at a specified angle.
After a full 90-degree rotation the $\lambda/2$ waveplate, the angle corresponding to the intensity maximum $\theta$, was recorded.
Next, the action of the two waveplates was swapped, i.e., the $\lambda/2$ waveplate remained fixed at the angle ($\theta$) and the $\lambda/4$ waveplate was rotated 90 degrees.
Again, the angle corresponding to the intensity maximum ($\phi$) was recorded.
Finally, the action of the waveplates was again swapped, with the $\lambda/4$ remaining fixed at $\phi$ and the $\lambda/2$ waveplate rotating.
This iterative process was then repeated until the angles at maximum intensity ($\theta, \phi$) no longer change after a few repetitions of the procedure.
This was done for both the exciton and biexciton emission wavelengths.
The end result of this calibration procedure was four waveplate angles, i.e., one for each waveplate on both the exciton and biexciton path, which corresponded to the waveplate settings for projecting onto the $H$ polarization state.
The waveplate setting (angles) for projective measurements on all other polarization states ($V, A, D, R, L$) are simply the $H$ settings shifted by some calculable amount \cite{James2001-vw}.
Assuming perfect waveplates, with the equation
\begin{equation}
    \ket{\psi_{proj}} = \hat{U}_{\lambda/4}(\phi) \hat{U}_{\lambda/2}(\theta) \ket{H} = 
    \frac{1}{2}
    \begin{pmatrix}
        \cos(\phi)^2 + i\sin(\phi)^2 & \frac{(1-i)}{2}\sin(2 \phi) \\
        \frac{(1-i)}{2}\sin(2 \phi) &  \sin(\phi)^2 + i\cos(\phi)^2
    \end{pmatrix}
    \begin{pmatrix}
        \cos(2 \theta) & \sin(2 \theta) \\
        \sin(2 \theta) & -\cos(2 \theta) 
    \end{pmatrix}
    \begin{pmatrix}
        1 \\
        0
    \end{pmatrix}
\end{equation}
we determine the angles $\theta$, $\phi$ required to implement a projective measurement along the state $\ket{\psi_{proj}}$, given that the waveplates are calibrated to $\ket{H} = \begin{pmatrix} 1 & 0\end{pmatrix}^{\intercal}$.
The angles we used are listed below in Table \ref{table:waveplate-settings}.
We verified the waveplate settings by performing a full 36-measurement quantum state tomography trial on the $H$-polarized laser.
With these calibrated settings, the system reconstructed the $H$-polarized laser input to a highly-pure state with a purity Tr$\left( \rho^2 \right) = 99.8 \pm 0.2 \%$.
The density matrix of the reconstructed state is plotted in Figure \ref{fig:calibration_dm}.

\begin{table}[!h]
\centering
\begin{tabular}{c|c|c|c|c|c|c}
$\ket{\psi_{proj}}$ & H & V  &  A    &  D    &  R   & L    \\ \hline
$\phi$ (QWP)        & 0 & 0  &  45   & 45    &  90  & 0    \\ \hline
$\theta$ (HWP)      & 0 & 45 & -22.5 & 22.5  & 22.5 & 22.5
\end{tabular}
\caption{\label{table:waveplate-settings} Waveplate angle settings used in the quantum state tomography experiments. }
\end{table}

\clearpage 
\renewcommand{\bibsetup}{ \thispagestyle{empty} } 
\printbibliography[heading=subbibliography]

\end{refsection}

\end{document}